\documentclass[twocolumn,showpacs,showkeys,preprintnumbers,amssymb,aps,superscriptaddress,prb]{revtex4}
\usepackage{graphicx}
\usepackage{dcolumn}
\usepackage{bm}
\usepackage{color}
\usepackage{epstopdf}
\usepackage[T1]{fontenc}
\begin{document}


\title{Breakdown of intermediate one-half magnetization plateau of spin-1/2 Ising-Heisenberg and Heisenberg branched chains at triple and Kosterlitz-Thouless critical points}
\author{Katar\'ina Kar{l}'ov\'a}
\email{katarina.karlova@student.upjs.sk}
\affiliation{Institute of Physics, Faculty of Science, P. J. \v{S}af\'{a}rik University, Park Angelinum 9, 04001 Ko\v{s}ice, Slovakia}
\author{Jozef Stre\v{c}ka}
\affiliation{Institute of Physics, Faculty of Science, P. J. \v{S}af\'{a}rik University, Park Angelinum 9, 04001 Ko\v{s}ice, Slovakia}
\author{Marcelo L. Lyra}
\affiliation{Instituto de Fisica, Universidade Federal de Alagoas, 57072-970 Macei\'o, AL, Brazil}

\date{\today}

\begin{abstract}
The spin-1/2 Ising-Heisenberg branched chain composed of regularly alternating Ising spins and Heisenberg dimers involving an additional side branching is rigorously solved in a magnetic field by the transfer-matrix approach. The ground-state phase diagram, the magnetization process and the concurrence measuring a degree of bipartite entanglement within the Heisenberg dimers are examined in detail.  Three different ground states were found depending on a mutual interplay between the magnetic field and two different coupling constants: the modulated quantum antiferromagnetic phase, the quantum ferrimagnetic phase and the classical ferromagnetic phase. Two former quantum ground states are manifested in zero-temperature magnetization curves as intermediate plateaus at zero and one-half of the saturation magnetization, whereas the one-half plateau disappears at a triple point induced by a strong enough ferromagnetic Ising coupling. The ground-state phase diagram and zero-temperature magnetization curves of the analogous spin-1/2 Heisenberg branched chain were investigated using DMRG calculations. The latter fully quantum Heisenberg model involves, besides two gapful phases  manifested as zero and one-half magnetization plateaus, gapless quantum spin-liquid phase. The intermediate one-half plateau of the spin-1/2 Heisenberg branched chain vanishes at Kosterlitz-Thouless quantum critical point between gapful and gapless quantum ground states unlike the triple point of the spin-1/2 Ising-Heisenberg branched chain.
\end{abstract}
\pacs{75.10.Jm, 75.30.Kz, 75.40.Cx, 03.65.Ud}
\keywords{Ising-Heisenberg model, Hesienberg model, branched chain, magnetization plateaus, DMRG simulations}

\maketitle

\section{Introduction}
For many years one-dimensional quantum spin chains have become of great scientific interest both from the theoretical point of view \cite{kura98,yama99,saka99,ivan00,hone00,yama00,saka02,teno11,karldi} as well as from the experimental point of view.\cite{mill01,hagi98,kopi82,kami05,kuhn09,jeon15,chak17,maes18} Quantum Heisenberg spin chains can for instance exhibit many interesting features in a magnetization process such as magnetization plateau,\cite{maru18,mori18} quantum spin liquid,\cite{lswu19,avalar,silv17} or quasiplateau.\cite{bell14,ohan15} Among these magnetization anomalies, the intermediate plateaus should obey a quantization condition known as Oshikawa-Yamanaka-Affleck rule $S_t-m_t={\rm{integer}}$, where $S_t$ and $m_t$ is a total spin and total magnetization per elementary unit, respectively.\cite{oshi97} 

\begin{figure*}[t]
\begin{center}
\includegraphics[width=0.7\textwidth]{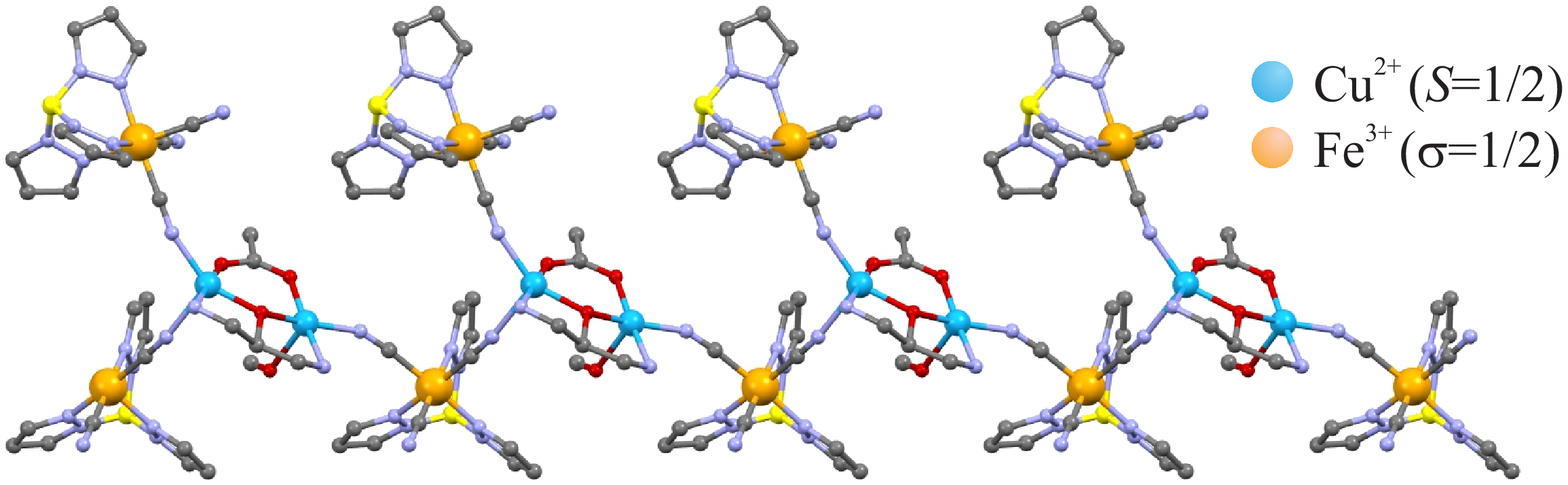}
\end{center}
\vspace{-0.8cm}
\caption{A part of the crystal structure of the heterobimetalic coordination polymer [(Tp)$_2$Fe$_2$(CN)$_6$(OAc)(bap)Cu$_2$(CH$_3$OH)$\cdot$2CH$_3$OH.H$_2$O] adopted according to the crystallographic data reported in Ref. \onlinecite{vetve}  Smaller (blue) circles determine positions of the divalent Cu$^{2^+}$ magnetic ions and larger (orange) circles determine lattice position of trivalent Fe$^{3^+}$ magnetic ions.}
\label{fig1}
\end{figure*}

Quantum ground states of the Heisenberg spin chains ensue and/or break down at quantum phase transitions, which may be however very different in character.\cite{karlar,veri19,ejim18,gliu15,saka99} For instance, the intermediate 3/7-plateau of the mixed spin-(1/2,5/2,1/2) Heisenberg branched chain terminates at the Kosterlitz-Thouless quantum critical point.\cite{veri19} In the present paper we will examine the ground-state phase diagram and magnetization process of the related spin-1/2 Ising-Heisenberg and Heisenberg branched chains, whose magnetic structure is inspired by the heterobimetallic coordination polymer [(Tp)$_2$Fe$_2$(CN)$_6$(OCH$_3$)(bap)Cu$_2$(CH$_3$OH)$\cdot$2CH$_3$OH.H$_2$O] (Tp=tris(pyrazolyl)hydroborate, bapH = 1,3-bis(amino)-2-propanol) \cite{vetve} to be further abbreviated as Fe$_2$Cu$_2$, which incorporates the highly anisotropic trivalent Fe$^{3+}$ cations and the almost isotropic divalent Cu$^{2+}$ cations (see Fig. \ref{fig1}). The magnetic features of the polymeric coordination compound Fe$_2$Cu$_2$ should be described within the framework of the spin-1/2 XXZ Heisenberg branched chain, which cannot be solved exactly. However, the trivalent Fe$^{3+}$ magnetic ions in a low-spin state ($S$=1/2) posses a relatively high degree of the magnetic anisotropy due to unquenched  orbital momentum. \cite{jong74,carl86,kahn93,wolf00} In this regard we will at first examine the spin-1/2 Ising-Heisenberg branched chain by considering the highly anisotropic trivalent Fe$^{3+}$ ions as the classical Ising spins and the almost isotropic divalent Cu$^{2+}$ cations as the quantum Heisenberg spins. This simplification allows a derivation of exact results for the spin-1/2 Ising-Heisenberg branched chain and often might be regarded as a good approximation of its full quantum XXZ Heisenberg counterpart.\cite{stre05,huev10,saho12,stre12,bell14,souz19,torr18,verk16} Beside this, we will adapt density-matrix renormalization group (DMRG) method in order to determine the ground-state phase diagram of the fully quantum spin-1/2 Heisenberg branched chain, which will be confronted with exact results for the simpler spin-1/2 Ising-Heisenberg branched chain. 

The organization of this paper is as follows. The spin-1/2 Ising-Heisenberg branched chain is introduced and solved by the use of the transfer-matrix method in Sec. \ref{sec:method}. The ground-state phase diagram, magnetization curves and the concurrence between the Heisenberg dimers of the spin-1/2 Ising-Heisenberg branched chain are presented in Sec. \ref{IH}. The most interesting results for the spin-1/2 Heisenberg branched chain are presented in Sec. \ref{H} and finally some summarized ideas are posted in Sec. \ref{conclusion}.

\section{Ising-Heisenberg branched chain}
\label{sec:method}

\begin{figure*}
\begin{center}
\includegraphics[width=0.55\textwidth]{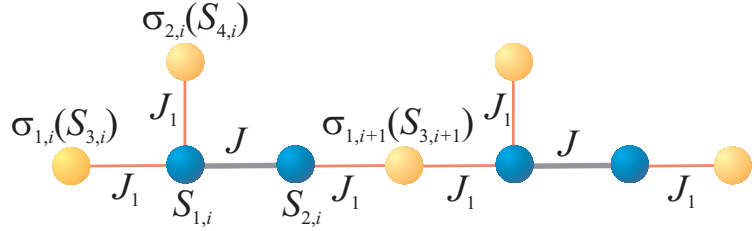}
\end{center}
\vspace{-0.4cm}
\caption{A schematic illustration of the spin-1/2 Ising-Heisenberg branched chain. Dark (blue) circles denote lattice positions of the Heisenberg spins and light (orange) circles denote  lattice positions of the Ising spins. The notation for an analogous pure spin-1/2 Heisenberg branched chain is given in round brackets whenever it differs from the spin-1/2 Ising-Heisenberg branched chain.}
\label{fig2}
\end{figure*}

Let us consider the spin-1/2 Ising-Heisenberg branched chain schematically depicted in Fig. \ref{fig2}, which can be defined through the following Hamiltonian
\begin{eqnarray} 
\label{ham}
\hat{\cal{H}}\!\!\!&=&\!\!\! \sum_{i=1}^N \left\{J\left[\Delta\left(\hat{S}_{1,i}^x\hat{S}_{2,i}^x + \hat{S}_{1,i}^y\hat{S}_{2,i}^y \right)
+ \hat{S}_{1,i}^z\hat{S}_{2,i}^z\right]\right. \nonumber \\ \!\!\!&-&\!\!\! J_1\left(\hat{S}_{1,i}^z\hat{\sigma}_{1,i}^z+\hat{S}_{2,i}^z\hat{\sigma}_{1,i+1}^z+\hat{S}_{1,i}^z\hat{\sigma}_{2,i}^z\right) \nonumber \\ \!\!\!&-&\!\!\! \left.  h_1\left(\hat{S}_{1,i}^z+\hat{S}_{2,i}^z\right) \!-h_2\hat{\sigma}_{1,i}^z\!-h_3\hat{\sigma}_{2,i}^z \right\}\!\!,
\end{eqnarray}
where $\hat{\sigma}_i^z$ and $\hat{S}_{i}^{\alpha}$ ($\alpha=x,y,z$) denote the spatial components of the spin-1/2 operators related to the Ising and Heisenberg spins, respectively.  The coupling constant $J>0$ stands for the antiferromagnetic Heisenberg interaction inside of dimeric Cu$^{2+}$-Cu$^{2+}$ units from a backbone of the polymeric chain, while the coupling constant $J_1>0$ ($J_1<0$) describes the ferromagnetic (antiferromagnetic) Ising-type interaction between nearest-neighbor Ising and Heisenberg spins approximating the magnetically anisotropic Fe$^{3+}$ and magnetically isotropic Cu$^{2+}$ ions, respectively. Last but not least, Zeeman's terms $h_j$ ($j=1,2,3$) are assigned to a coupling of the Ising and Heisenberg spins with an external magnetic field,  $N$ denotes the total number of unit cells. For simplicity, periodic boundary conditions $\sigma_{1,N+1}\equiv\sigma_{1,1}$ are assumed. 

For further convenience it is advisable  to rewrite the Hamiltonian (\ref{ham}) as a sum of the cell Hamiltonians
\begin{eqnarray} 
\hat{\cal{H}}= \sum_{i=1}^N \hat{\cal{H}}_i,
\label{sum}
\end{eqnarray}
where the cell Hamiltonian $\hat{\cal{H}}_i$ is defined by 
\begin{eqnarray} 
\label{hami}
\hat{\cal{H}}_i\!\!\!&=&\!\!\! J\left[\Delta\left(\hat{S}_{1,i}^x\hat{S}_{2,i}^x + \hat{S}_{1,i}^y\hat{S}_{2,i}^y \right)
+ \hat{S}_{1,i}^z\hat{S}_{2,i}^z\right]\nonumber \\ \!\!\!&-&\!\!\! J_1\left(\hat{S}_{1,i}^z\hat{\sigma}_{1,i}^z+\hat{S}_{2,i}^z\hat{\sigma}_{1,i+1}^z+\hat{S}_{1,i}^z\hat{\sigma}_{2,i}^z\right) \nonumber \\ \!\!\!&-&\!\!\!  h_1\left(\hat{S}_{1,i}^z+\hat{S}_{2,i}^z\right)-h_3\hat{\sigma}_{2,i}^z -\frac{h_2}{2}\left(\hat{\sigma}_{1,i}^z + \hat{\sigma}_{1,i+1}^z\right)\!\!.
\end{eqnarray}
The cell Hamiltonians $\hat{\cal{H}}_i$ apparently commute, i.e. $[\hat{\cal{H}}_i,\hat{\cal{H}}_j]=0$, which means that the partition function of the spin-1/2  Ising-Heisenberg branched chain can be written in this form
\begin{eqnarray} 
\label{PF}
\cal{Z} \!\!\!&=&\!\!\! \sum_{\{\sigma\}}{\rm{Tr}}~{\rm{e}}^{-\beta\sum_i\hat{\cal{H}}_i}=\sum_{\{\sigma_{1,i}^z\}}\prod_{i=1}^N \sum_{\sigma_{2,i}^z}{\rm{Tr}}_{[S_{1,i}, S_{2,i}]}{\rm{e}}^{-\beta \hat{\cal{H}}_i}\nonumber \\
 \!\!\!&=&\!\!\! \sum_{\{\sigma_{1,i}^z\}}\prod_{i=1}^N {\rm{T}}(\sigma_{1,i}^z;\sigma_{1;i+1}^z),
\end{eqnarray}
where $\beta=1/(k_{\rm{B}}T)$, $k_{\rm{B}}$ is the Boltzmann's factor and $T$ is the absolute temperature. ${\rm{Tr}}_{[S_{1,i}, S_{2,i}]}$ denotes a trace over degrees of the Heisenberg dimer from the $i$-th unit cell, the symbol $\sum_{\{\sigma_{1,i}^z\}}$ marks a summation over all possible spin configurations of the  Ising spins from a backbone chain and the expression
\begin{eqnarray} 
\label{trace}
{\rm{T}}(\sigma_{1,i}^z;\sigma_{1;i+1}^z)= \sum_{\sigma_{2,i}^z}{\rm{Tr}}_{[S_{1,i}, S_{2,i}]}{\rm{e}}^{-\beta \hat{\cal{H}}_i}
\end{eqnarray}
is the effective Boltzmann's factor obtained after tracing out spin degrees of freedom of two Heisenberg spins and the Ising spin $\sigma_{2,i}$ at lateral branching. To proceed further with a calculation, one necessarily needs to evaluate the effective Boltzmann's factor given by Eq. (\ref{trace}). For this purpose, it is advisable to pass to a matrix representation of the cell Hamiltonian $\hat{\cal{H}}_i$ in the basis spanned over four available states of two Heisenberg spins $S_{1,i}$ and $S_{2,i}$
\begin{eqnarray} 
\label{baza}
|\!\uparrow,\uparrow\rangle_i=|\!\uparrow\rangle_{1,i}|\!\uparrow\rangle_{2,i}, \quad |\!\uparrow,\downarrow\rangle_i=|\!\uparrow\rangle_{1,i}|\!\downarrow\rangle_{2,i}, \nonumber \\ |\!\downarrow,\uparrow\rangle_i=|\!\downarrow\rangle_{1,i}|\!\uparrow\rangle_{2,i}, \quad |\!\downarrow,\downarrow\rangle_i=|\!\downarrow\rangle_{1,i}|\!\downarrow\rangle_{2,i}.
\end{eqnarray}
whereas $|\!\uparrow\rangle_{k,i}$ and $|\!\downarrow\rangle_{k,i}$ ($k=1,2$) denote two eigenvectors of the spin operator $\hat{S}_{k,i}^z$ with the respective eigenvalues $S_{k,i}^z = \pm 1/2$. After a straightforward diagonalization of the cell Hamiltonian $\hat{\cal{H}}_i$ one obtains the following four eigenvalues
\begin{eqnarray} 
\label{vlastneHodnoty}
E_{1i,2i} \!&=&\! \frac{J}{4} \pm \frac{J_1}{2}\left(\sigma_{1,i}^z+\sigma_{2,i}^z + \sigma_{1,i+1}^z\right) \pm h_1 \nonumber \\ \!\!\!\!&-&\!\!\!\! \frac{h_2}{2}\left(\sigma_{1,i}^z+ \sigma_{1,i+1}^z\right)-h_3\sigma_{2,i}^z, \nonumber \\
E_{3i,4i} \!&=&\! -\frac{J}{4} \pm \frac{1}{2}\sqrt{\left(\sigma_{1,i}^z\!+\!\sigma_{2,i}^z - \sigma_{1,i+1}^z\right)^2 \!+\!(J\Delta)^2} \nonumber \\ \!\!\!\!&-&\!\!\!\! \frac{h_2}{2}\left(\sigma_{1,i}^z+ \sigma_{1,i+1}^z\right)-h_3\sigma_{2,i}^z 
\end{eqnarray}
and the corresponding eigenvectors
\begin{eqnarray} 
\label{vlastnevectorky}
&&|\varphi_{1,i}\rangle=|\uparrow\rangle_{1,i}|\uparrow\rangle_{2,i}, \nonumber \\
&&|\varphi_{2,i}\rangle=|\downarrow\rangle_{1,i}|\downarrow\rangle_{2,i}, \nonumber \\
&&|\varphi_{3,i}\rangle= c_+|\uparrow\rangle_{1,i}|\downarrow\rangle_{2,i} + c_-|\uparrow\rangle_{1,i}|\downarrow\rangle_{2,i}, \nonumber \\
&&|\varphi_{4,i}\rangle= c_+|\uparrow\rangle_{1,i}|\downarrow\rangle_{2,i} - c_-|\uparrow\rangle_{1,i}|\downarrow\rangle_{2,i},
\end{eqnarray}
where 
\begin{eqnarray} 
\label{vlastnekoef}
c_{\pm}=\frac{1}{\sqrt{2}}\sqrt{1\pm\frac{J_1\left(\sigma_{1,i+1}^z-\sigma_{1,i}^z-\sigma_{2,i}^z\right)}{\sqrt{J_1^2\left(\sigma_{1,i+1}^z-\sigma_{1,i}^z-\sigma_{2,i}^z\right)^2+\left(J\Delta\right)^2}}}.
\end{eqnarray}
Note that the effective Boltzmann's factor ${\rm{T}}(\sigma_{1,i}^z;\sigma_{1;i+1}^z)$ given by Eq. (\ref{trace}) depends only on two Ising spins from backbone of the spin chain and can be alternatively viewed as the transfer matrix defined by
\begin{eqnarray} 
\label{trace2}
&&\!\!\!\!\!\!\!\!\!\!{\rm{T}}(\sigma_{1,i}^z;\sigma_{1;i+1}^z)\!\!=\!\!\! \sum_{\sigma_{2,i}^z}\!{\rm{Tr}}_{[S_{1,i}, S_{2,i}]}{\rm{e}}^{-\beta \hat{\cal{H}}_i}\!\!=\!\!\!\!\!\!\sum_{\sigma_{2,i}^z=\pm \frac{1}{2}}\sum_{j=1}^4{\rm{e}}^{-\beta E_{ji}}\nonumber \\
\!&=&\! 2{\rm{e}}^{\frac{\beta h_3}{2}\left(\sigma_{1,i}^z+\sigma_{1,i+1}^z\right)-\frac{\beta J}{4}}  \nonumber \\
\!\!\!&\times&\!\!\! \left\{{\rm{e}}^\frac{\beta  h_2}{2}\cosh\left[\frac{\beta}{2}\left(J_1\sigma_{1,i}^z+J_1\sigma_{1,i+1}^z +\frac{J_1}{2}+2h_1\right)\right] \right.\nonumber \\ 
\!\!\!&+&\!\!\! {\rm{e}}^{-\frac{\beta h_2}{2}}\cosh\left[\frac{\beta}{2}\left(J_1\sigma_{1,i}^z+J_1\sigma_{1,i+1}^z -\frac{J_1}{2}+2h_1\right)\right] \nonumber \\
\!\!\!&+&\!\!\!{\rm{e}}^{\frac{\beta J}{2}+\frac{\beta h_2}{2}}\!\!\cosh\!\!\left[\frac{\beta}{2}\sqrt{\!\!\left(J_1\sigma_{1,i}^z\!-\!J_1\sigma_{1,i+1}^z \!+\!\frac{J_1}{2}\right)^2 \!\!\!+\! (J\Delta)^2}\right] \nonumber \\
\!\!\!&+&\!\!\!\left.{\rm{e}}^{\frac{\beta J}{2}-\frac{\beta h_2}{2}}\!\!\cosh\!\!\left[\frac{\beta}{2}\sqrt{\!\!\left(J_1\sigma_{1,i}^z\!-\!J_1\sigma_{1,i+1}^z \!-\!\frac{J_1}{2}\right)^2 \!\!\!+\! (J\Delta)^2}\right]\!\! \right\}\!. \nonumber \\ 
\end{eqnarray}
A successive summation over states of a set of the Ising spins $\{\sigma_{1,i}\}$ from the backbone of a spin chain gives within the standard transfer-matrix approach \cite{transfer} the following final formula for the partition function 
\begin{eqnarray} 
\label{Zfinal}
{\cal{Z}}=\sum_{\{\sigma_{1,i}\}}\prod_{i=1}^N {\rm{T}}(\sigma_{1,i}^z;\sigma_{1;i+1}^z) =  {\rm{Tr}}~{\rm{T}}^N=\lambda_+^N+\lambda_-^N,
\end{eqnarray}
which depends on two eigenvalues of the transfer matrix (\ref{trace2})

\begin{eqnarray} 
\label{lambdy}
\lambda_{\pm}=\frac{T_1+T_2}{2}\pm\sqrt{\left(\frac{T_1-T_2}{2}\right)^2+T_3T_4}.
\end{eqnarray}
Here, the expressions $T_i$ ($i=1,2,3,4$) denote specific elements of the transfer matrix  (\ref{trace2}) for the following particular spin states of the Ising spins $\sigma_{1,i}$ and $\sigma_{1,i+1}$
\begin{eqnarray} 
T_1={\rm{T}}\left(\frac{1}{2},\frac{1}{2}\right), \quad 
T_2={\rm{T}}\left(-\frac{1}{2},-\frac{1}{2}\right), \nonumber \\
T_3={\rm{T}}\left(\frac{1}{2},-\frac{1}{2}\right), \quad
T_4={\rm{T}}\left(-\frac{1}{2},\frac{1}{2}\right).
\end{eqnarray}

At this stage, the exact result for the partition function (\ref{Zfinal}) can be used to obtain the Gibbs free energy, which is given in the thermodynamic limit only by the largest transfer-matrix eigenvalue
\begin{eqnarray} 
\label{G}
G=-k_{\rm{B}}T\lim_{N\rightarrow\infty}\frac{1}{N}\ln{\cal{Z}}=-k_{\rm{B}}T\ln\lambda_+.
\end{eqnarray}
One can subsequently obtain local magnetizations (and consequently the total magnetization) by differentiating the Gibbs free energy (\ref{G}) with respect to local magnetic fields

\begin{eqnarray} 
m_{1,2}\!&=&\!\frac{1}{2}\langle \hat{S}_{1,i}^z + \hat{S}_{2,i}^z\rangle =-\frac{1}{2N}\frac{\partial G}{\partial h_1}, \nonumber \\
m_3\!&=&\!\frac{1}{2}\langle \hat{\sigma}_{1,i}^z + \hat{\sigma}_{1,i+1}^z\rangle = -\frac{1}{N}\frac{\partial G}{\partial h_2}, \nonumber \\
m_4\!&=&\!\langle \hat{\sigma}_{2,i}^z\rangle = -\frac{1}{N}\frac{\partial G}{\partial h_3}, \nonumber \\
m_t\!&=&\!\frac{1}{4} \left( 2m_1 + m_2 +m_3\right).
\label{magneti}
\end{eqnarray}

To bring insight into a degree of bipartite entanglement inside of the Heisenberg dimers one may take advantage of the concurrence, \cite{wooters} which can be expressed in terms of the local magnetization and the respective pair correlation functions. \cite{amico,horodecki} To this end,  one can perform relevant differentiation of the Gibbs free energy (\ref{G}) with respect to  spatial components of the coupling constant in order to calculate respective spatial components of the pair correlation function between the nearest-neighbor Heisenberg spins according to the formulas
\begin{eqnarray} 
C^{zz}\!&=&\!\langle \hat{S}_{1,i}^z\hat{S}_{2,i}^z \rangle = -\frac{\partial \ln {\cal{Z}}}{N \partial(\beta J)},\nonumber \\
C^{xx}\!&=&\!\langle \hat{S}_{1,i}^x\hat{S}_{2,i}^x \rangle = -\frac{\partial \ln {\cal{Z}}}{N \partial(\beta J \Delta)}. 
\label{korelacne}
\end{eqnarray}
The concurrence can be then calculated from  the exact results for the pair correlation functions (\ref{korelacne}) and the local magnetization (\ref{magneti}) according to the formula \cite{wooters,amico,horodecki}
\begin{eqnarray} 
C={\rm{max}} \left\{0; 4|C^{xx}|-2\sqrt{\left(\frac{1}{4}+C^{zz}\right)^2-m_1^2}\right\}.
\label{konkarenc}
\end{eqnarray}

\section{The most interesting results}
\label{IH}

\begin{figure*}[t]
\begin{center}
\includegraphics[width=0.4\textwidth]{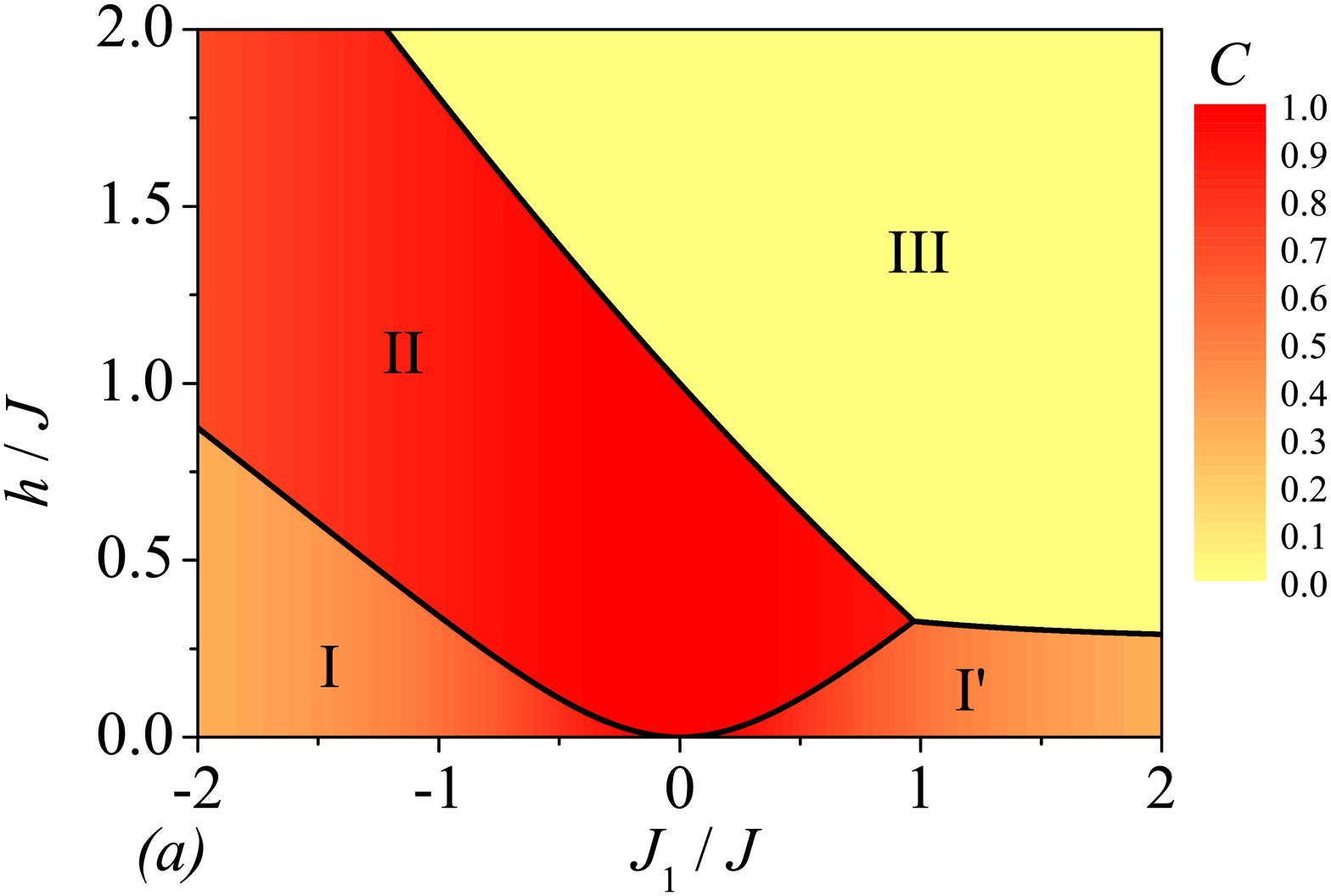}
\hspace{-0.2cm}
\includegraphics[width=0.4\textwidth]{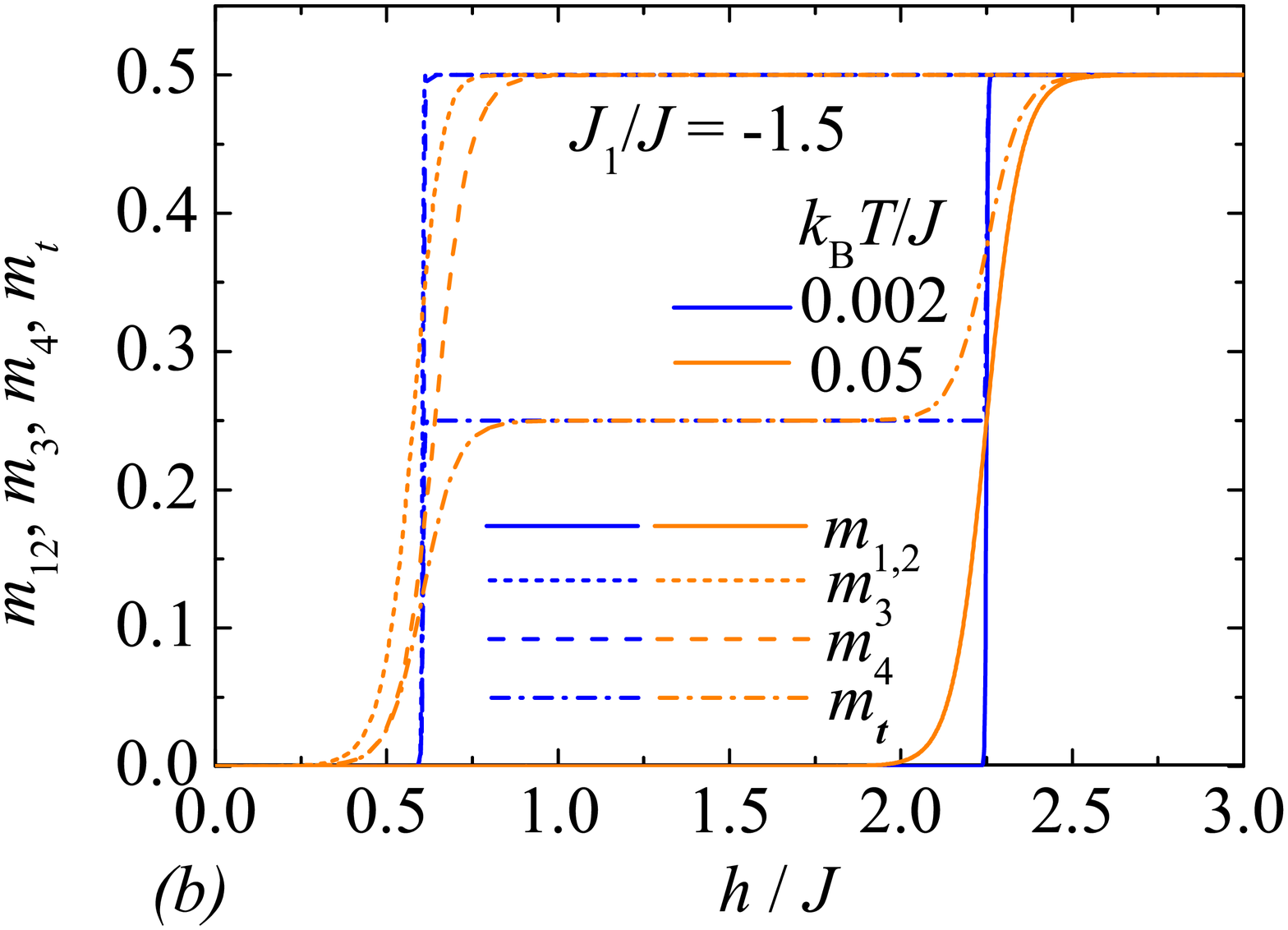}
\includegraphics[width=0.4\textwidth]{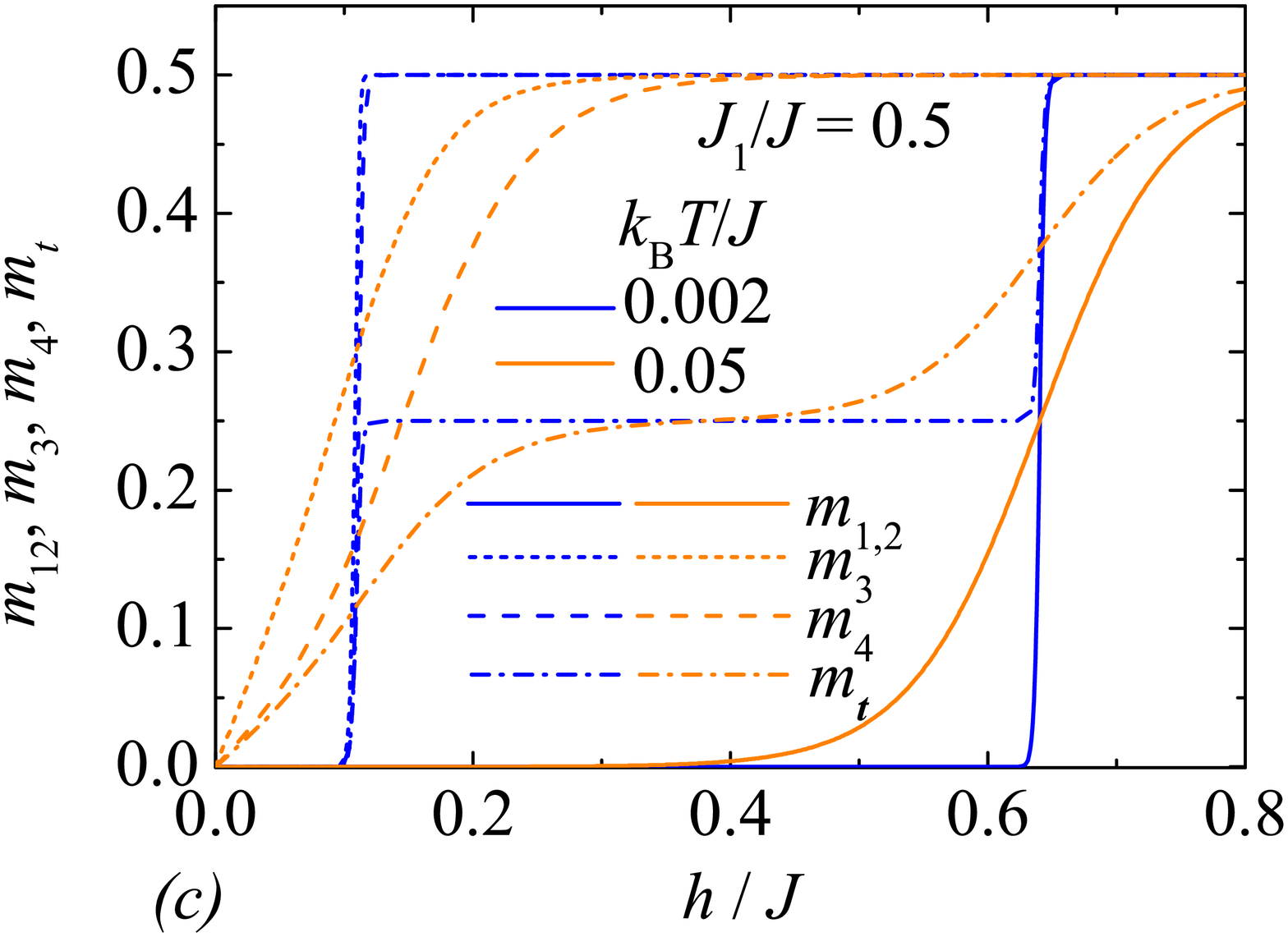}
\hspace{-0.2cm}
\includegraphics[width=0.4\textwidth]{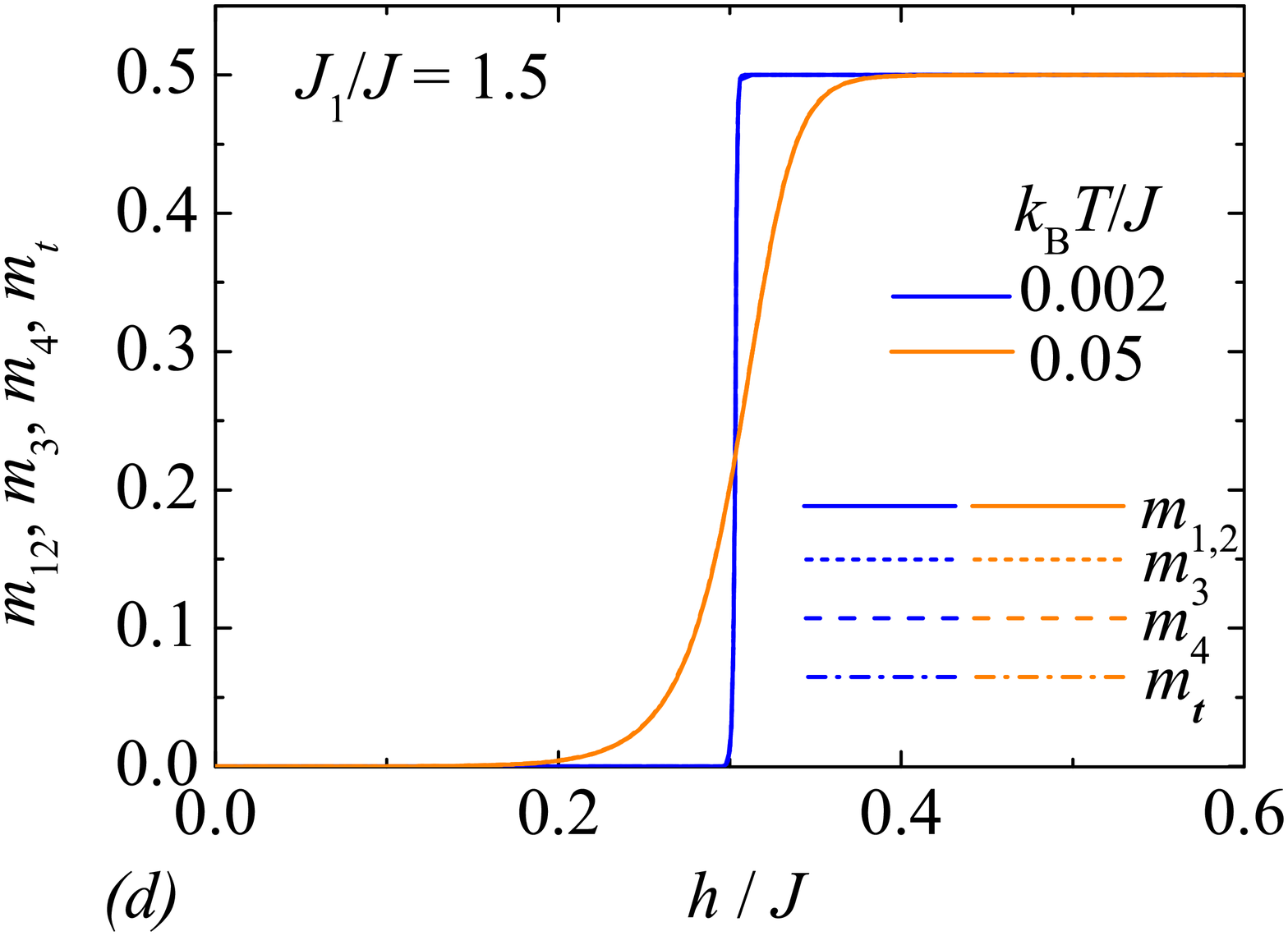}
\end{center}
\vspace{-0.4cm}
\caption{(a) The ground-state phase diagram of the spin-1/2 Ising-Heisenberg branched chain in the $J_1/J-h/J$ plane supplemented with a density plot of the concurrence calculated for the Heisenberg dimers; (b)-(d) magnetic-field dependencies of local and total magnetizations defined according to Eq. (\ref{magneti}) at two different temperatures and three values of the interaction ratio: (b) $J_1/J=-1.5$; (c) $J_1/J=0.5$; (d) $J_1/J=1.5$.}
\label{fig3}
\end{figure*}
\begin{figure}[t]
\begin{center}
\includegraphics[width=0.45\textwidth]{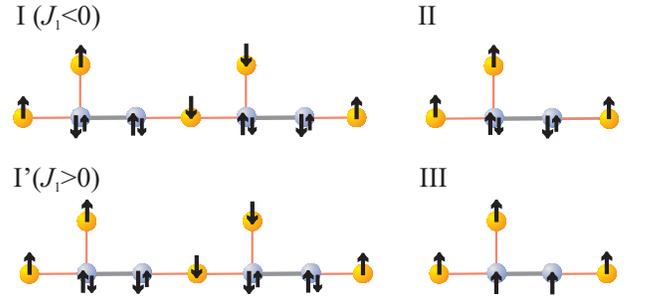}
\end{center}
\vspace{-0.4cm}
\caption{A schematic illustration of spin arrangments within all possible ground states of the spin-1/2 Ising-Heisenberg branched chain. Two arrows within a single site denote quantum superposition of both spin states, whereas larger (smaller) arrow refers to a spin state emergent with a greater (smaller) occurrence probability.}
\label{fig4}
\end{figure}

 In what follows, we will consider the particular case of the spin-1/2 Ising-Heisenberg branched chain  with the isotropic ($\Delta = 1$) antiferromagnetic Heisenberg interaction $J>0$, which will henceforth serve as an energy unit when defining dimensionless interaction parameters $J_1/J$ and $h/J$ measuring a relative strength of the coupling constants and magnetic field, respectively. For simplicity, we will further assume that all particular local magnetic fields are the same $h=h_1=h_2=h_3$ what corresponds to assuming equal Land\'e $g$-factors of Cu$^{2+}$ and Fe$^{3+}$ magnetic ions. By comparing energies of all lowest-energy  eigenstates one can obtain the ground-state phase diagram of the spin-1/2 Ising-Heisenberg branched chain as depicted in Fig. \ref{fig3}(a) in $J_1/J-h/J$ plane. Solid lines in Fig. \ref{fig3}(a) denote discontinuous field-induced phase transitions, which split the overall parameter space into three regions labeled as I (I'), II and III. The microscopic character of the relevant ground states is schematically shown in Fig. \ref{fig4} and the corresponding eigenvectors are given by
\begin{widetext}
\begin{eqnarray}
|{\rm{I,I'}}\rangle \!\!&=&\!\! \prod_{i=1}^{N/2}\!\!\left\{\!\left[a_+|\!\downarrow\rangle_{S_{1,2i-1}}|\!\uparrow\rangle_{S_{2,2i-1}}\!\!-a_-|\!\uparrow\rangle_{S_{1,2i-1}}|\!\downarrow\rangle_{S_{2,2i-1}}\right] |\!\!\uparrow\rangle_{\sigma_{1,2i-1}}|\!\uparrow\rangle_{\sigma_{2,2i-1}} \right.\nonumber \\
\!\!&+&\!\! \left. \left[(a_-|\!\downarrow\rangle_{S_{1,2i}}|\!\uparrow\rangle_{S_{2,2i}}\!-a_+|\!\uparrow\rangle_{S_{1,2i}}|\!\downarrow\rangle_{S_{2,2i}}\right]\!|\!\downarrow\rangle_{\sigma_{1,2i}}|\!\downarrow\rangle_{\sigma_{2,2_i}}\right\}\!\!, \nonumber \\
{|\rm{II}}\rangle \!\!&=&\!\! \prod_{i=1}^{N}\left[b_+|\!\downarrow\rangle_{S_{1,i}}|\!\uparrow\rangle_{S_{2,i}}-b_-|\!\uparrow\rangle_{S_{1,i}}|\downarrow\rangle_{S_{2,i}}\right]|\!\uparrow\rangle_{\sigma_{1,i}}|\!\uparrow\rangle_{\sigma_{2,i}},\nonumber\\
|{\rm{III}}\rangle \!\!&=&\!\! \prod_{i=1}^{N}|\!\uparrow\rangle_{S_{1,i}}|\!\uparrow\rangle_{S_{2,i}}|\!\uparrow\rangle_{\sigma_{1,i}} |\!\uparrow\rangle_{\sigma_{2,i}}.
\label{fazicky}
\end{eqnarray}
\end{widetext}
The respective probability amplitudes are defined as
\begin{eqnarray}
a_{\pm} = \frac{1}{\sqrt{2}}\sqrt{1\pm\frac{\frac{3J_1}{2}}{\sqrt{\left(\frac{3J_1}{2}\right)^2 + (J\Delta)^2}}} 
\end{eqnarray}
and 
\begin{eqnarray}
b_{\pm} = \frac{1}{\sqrt{2}}\sqrt{1\pm\frac{\frac{J_1}{2}}{\sqrt{\left(\frac{J_1}{2}\right)^2 + (J\Delta)^2}}}. 
\end{eqnarray}
The ground state  I (I') can be viewed as the modulated antiferromagnetic phase with a twofold  breaking of translational symmetry, which involves a singlet-like state of the Heisenberg dimers and up-up-down-down spin arrangements of the Ising spins. Note furthermore that the ground states $|{\rm{I}}\rangle$ and $|{\rm{I'}}\rangle$ emergent for the antiferromagnetic ($J_1<0$) and ferromagnetic ($J_1>0$) Ising coupling differ from one another just by a relative orientation of the singlet-like state of the Heisenberg dimers with respect to its surrounding Ising spins. The ground state $|{\rm{II}}\rangle$ has character of the quantum ferrimagnetic phase with other singlet-like state of the Heisenberg dimers accompanied with the fully polarized Ising spins. It is noteworthy that these ground states have obvious quantum features as exemplified by nonzero concurrence serving as a measure of bipartite entanglement within the Heisenberg dimers. Finally, the third ground state $|{\rm{III}}\rangle$ is classical  ferromagnetic phase with fully polarized Ising as well as Heisenberg spins. The ground-state phase boundaries [solid lines in Fig. \ref{fig3}(a)] are given by the following exact prescriptions 
\begin{enumerate}
\item  phase boundary $|{\rm{I}}\rangle$($|{\rm{I'}}\rangle$)/$|{\rm{II}}\rangle$:
\begin{eqnarray} 
\frac{h}{J}=\frac{1}{2}\left[ \sqrt{\left(\frac{3}{2}\frac{J_1}{J}\right)^2+1}-\sqrt{\left(\frac{1}{2}\frac{J_1}{J}\right)^2+1}\right],
\end{eqnarray}
\item  phase boundary $|{\rm{I'}}\rangle$/$|{\rm{III}}\rangle$:
\begin{eqnarray} 
\label{f15}
\frac{h}{J}=\frac{1}{4}-\frac{3}{8}\frac{J_1}{J} + \frac{1}{4}\sqrt{\left(\frac{3}{2}\frac{J_1}{J}\right)^2+1},
\end{eqnarray}
\item  phase boundary  $|{\rm{II}}\rangle$/$|{\rm{III}}\rangle$:
\begin{eqnarray} 
\frac{h}{J}=\frac{1}{2}-\frac{3}{4}\frac{J_1}{J} + \frac{1}{2}\sqrt{\left(\frac{1}{2}\frac{J_1}{J}\right)^2+1}.
\end{eqnarray}
\end{enumerate}

The local and total magnetizations are plotted in Fig. \ref{fig3}(b) against the external magnetic field for the special case $J_1/J=-1.5$ and two different temperatures. The local magnetization $m_{1,2}$ defined as the mean magnetization of the Heisenberg dimers exhibits only zero magnetization plateau, because the Heisenberg dimers do not contribute to the total magnetization up to a saturation field due to their singlet-like character in the phases $|{\rm{I}}\rangle$ and $|{\rm{II}}\rangle$ (see Fig. \ref{fig4}). The local magnetizations $m_3$ and $m_4$, which refer to the magnetization of the Ising spins within the main chain ($m_3$) and side branching ($m_4$), respectively, are  zero only at low enough
 magnetic fields due to their up-up-down-down spin alignment realized within the phase $|{\rm{I}}\rangle$. 
However, the total magnetization exhibits a discontinuous  magnetization jump due to a field-driven phase transition from zero plateau (the phase $|{\rm{I}}\rangle$) towards the one-half plateau (the phase $|{\rm{II}}\rangle$), which relates to a reorientation of the Ising spins (i.e. the local magnetizatons $m_3$ and $m_4$) towards the magnetic field. It should be nevertheless remarked that at finite temperatures there is no true magnetization plateau and jump  neither in Fig. \ref{fig3}(b), nor in all other figures, because  the actual magnetization jump and plateau exist at zero temperature only. It is evident from Fig. \ref{fig3}(b) that 
the magnetization curve at low enough temperature $k_{\rm B}T/J=0.002$ shows a steep but continuous rise with the magnetic field, while  an increase of temperature generally causes a gradual smoothing of the magnetization curve.  The second discontinuous field-driven phase transition appears at a saturation field, at which a spin reorientation of the Heisenberg spins takes place.

The local and total magnetizations are depicted in Fig. \ref{fig3}(c) as a function of the magnetic field for another value of the interaction ratio $J_1/J=0.5$. In this parameter space the phases $|{\rm{I'}}\rangle$, $|{\rm{II}}\rangle$ and $|{\rm{III}}\rangle$ can be realized as the respective ground states depending on a relative size of the magnetic field. The only change with respect to the aforementioned particular case lies in a change of relative orientation of the nearest-neighbor Ising and Heisenberg spins, which relates to different character of the Ising coupling constant. At very low magnetic fields, the phase $|{\rm{I'}}\rangle$ is realized as the relevant ground state, whose microscopic character implies zero contribution of all local magnetizations to the total magnetization. Above the first critical field $h_c/J\approx 0.11$ the phase $|{\rm{II}}\rangle$ becomes the ground state with zero contribution of the local magnetization $m_1$ of the Heisenberg dimers and saturated values of the local magnetizations $m_3$ and $m_4$ of the Ising spins. It is obvious from Fig. \ref{fig3}(c) that temperature $k_{\rm B}T/J=0.05$ is high enough to destroy zero-magnetization plateau of the local magnetization $m_3$ and $m_4$. 

Last but not least, we have investigated the local and total magnetizations in the parameter space supporting only the phases $|{\rm{I'}}\rangle$ and $|{\rm{III}}\rangle$ as the respective ground states. To support this statement, the local and total magnetizations are plotted in Fig. \ref{fig3}(d) against the external magnetic field for the special case $J_1/J=1.5$. The phase  $|{\rm{I'}}\rangle$ is realized as the ground state below the critical field $h_{c}/J \approx 0.30$, while the   classical fully polarized ferromagnetic phase $|{\rm{III}}\rangle$ becomes the ground state above this critical field. All local magnetizations of the Ising and Heisenberg spins behave alike in this particular case and thus, they cannot be discerned within the displayed figure. 

\begin{figure*}[t]
\begin{center}
\includegraphics[width=0.45\textwidth]{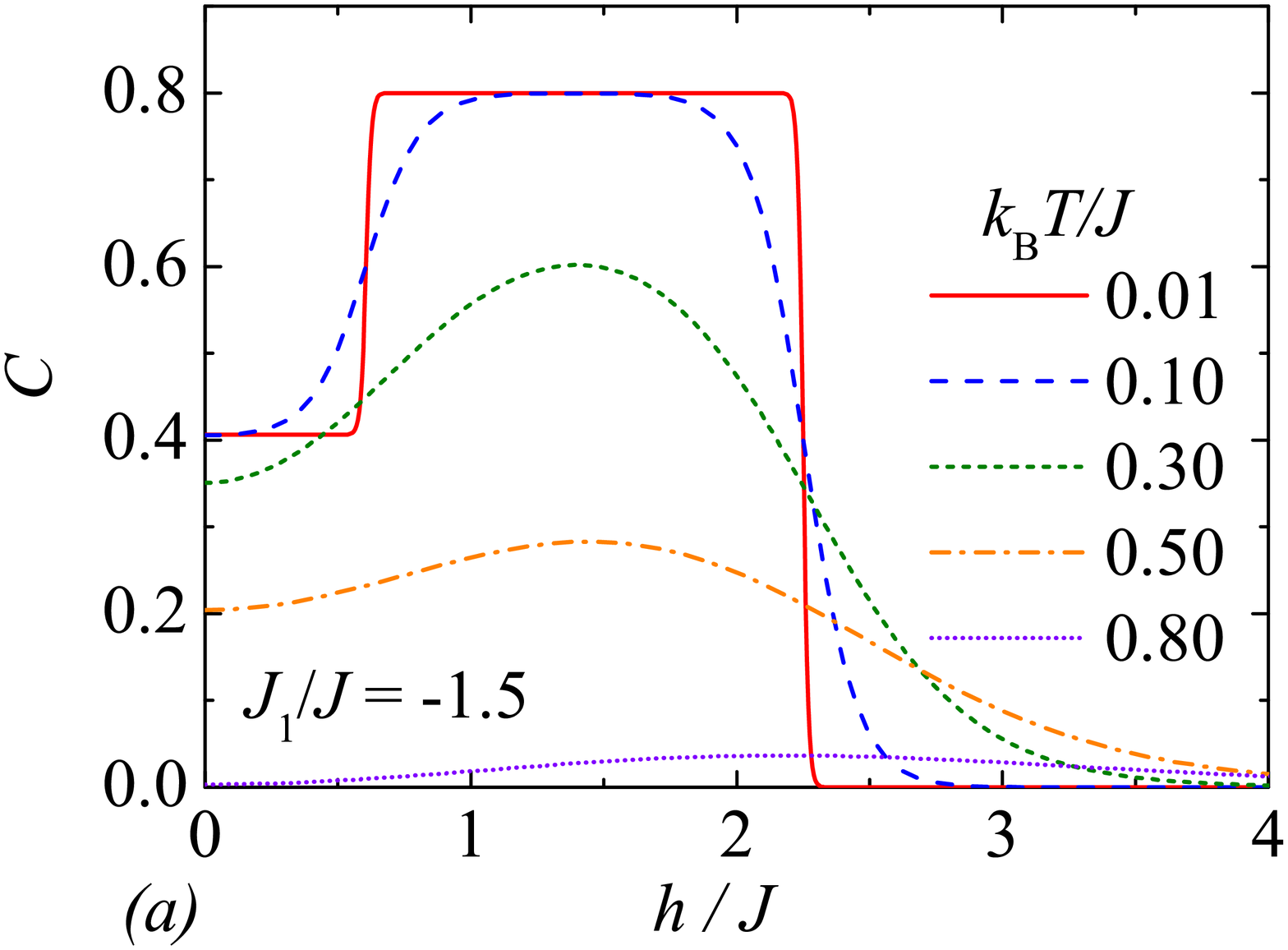}
\hspace{-0.2cm}
\includegraphics[width=0.45\textwidth]{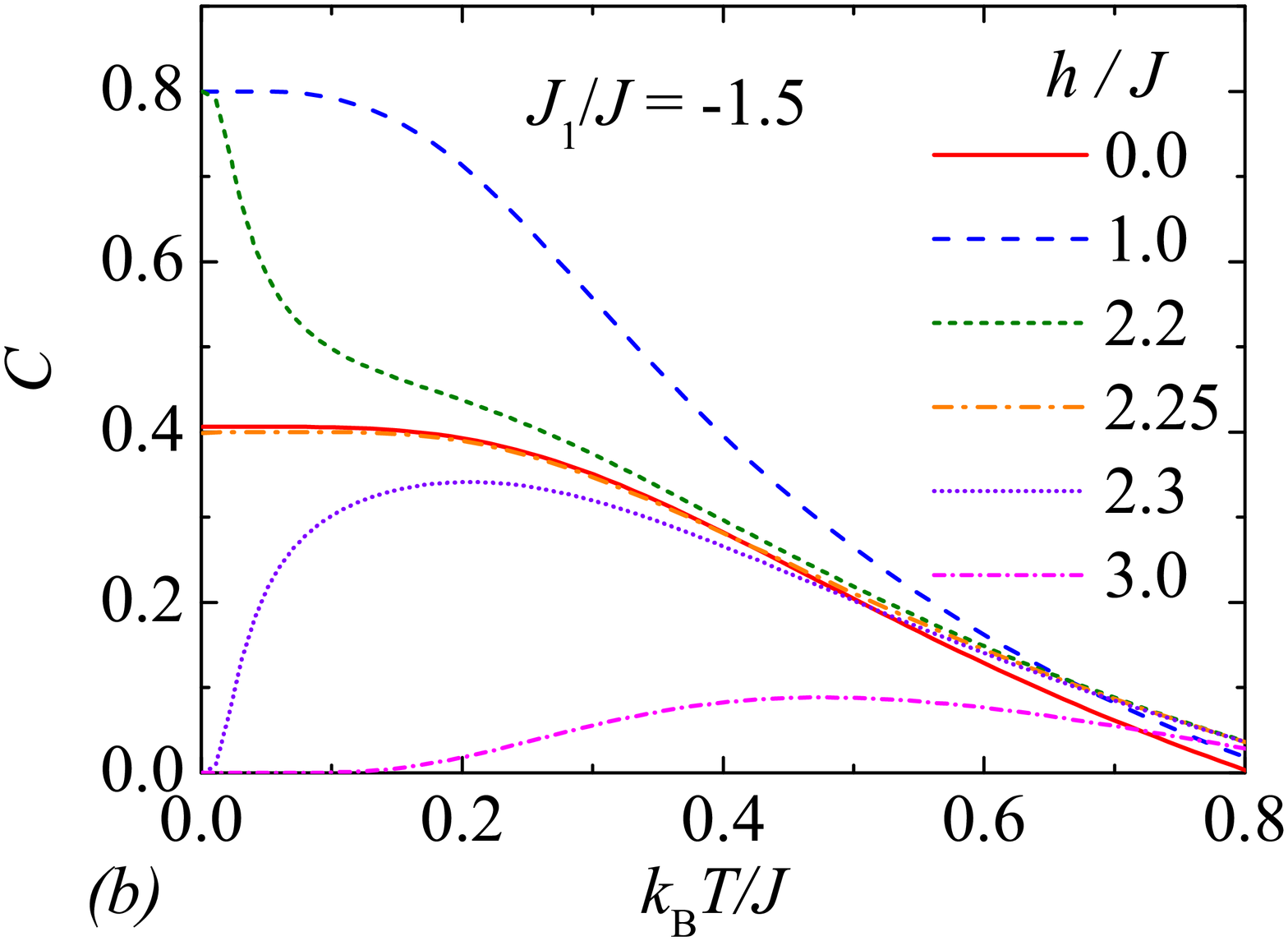}
\hspace{-0.2cm}
\end{center}
\vspace{-0.4cm}
\caption{(a) The concurrence as a function of the magnetic field for the interaction ratio $J_1/J=-1.5$ and several values of temperature; (b) Temperature variations of the concurrence for the interaction ratio $J_1/J=-1.5$ and several values of magnetic field.}
\label{fig5}
\end{figure*}
To bring a deeper  insight into a degree of bipartite entanglement between the nearest-neighbor Heisenberg spins (dimers) we will comprehensively examine the concurrence as a function of the magnetic field and temperature in three different cuts of the parameter space. 
 It is obvious from the eigenvectors (\ref{fazicky}) that the Heisenberg dimers are in the ground states $|{\rm{I}}\rangle$ ($|{\rm{I'}}\rangle$) and $|{\rm{II}}\rangle$ quantum-mechanically entangled, whereas the concurrence characterizing bipartite entanglement achieves in a zero-temperature limit the following asymptotic values
\begin{eqnarray} 
&& C_{\rm{I}}=C_{\rm{I'}}=\frac{1}{\sqrt{\left(\frac{3}{2}\frac{J_1}{J}\right)^2 + 1}}, \nonumber \\
&& C_{\rm{II}}= \frac{1}{\sqrt{\left(\frac{1}{2}\frac{J_1}{J}\right)^2 +1}}.
\label{conkifaz}
\end{eqnarray}
 It follows from the formulas (\ref{conkifaz}) that Heisenberg dimers are more strongly entangled in the ferrimagnetic phase $|{\rm{II}}\rangle$ than in the modulated antiferromagnetic phase $|{\rm{I}}\rangle$ ($|{\rm{I'}}\rangle$) for the same value of the interaction ratio $J_1/J$. 
The field dependence of the concurrence is displayed in Fig. \ref{fig5}(a) for the interaction ratio $J_1/J=-1.5$ and several values of temperature. The low-temperature asymptotes ($k_{\rm{B}}T/J=0.01$) of the  concurrence can be understood from the ground-state phase diagram [Fig. \ref{fig3}(a)] and the formulas (\ref{conkifaz}), which imply existence of three different ground states $|{\rm{I}}\rangle$, $|{\rm{II}}\rangle$ and $|{\rm{III}}\rangle$ depending on a relative size of the magnetic field. It follows from Fig. \ref{fig5}(a) that the concurrence is kept constant  at low enough temperatures and then it shows abrupt changes in a vicinity of the critical fields associated with the magnetization jumps. The non-zero value of the concurrence up to the second critical field $h_{c2}/J=2.25$ at sufficiently low temperatures ($k_{\rm{B}}T/J=0.01$) proves quantum character of the phases $|{\rm{I}}\rangle$ and $|{\rm{II}}\rangle$, while the zero concurrence at higher magnetic fields $h/J>2.25$ confirms classical character of the phase $|{\rm{III}}\rangle$. Interestingly, the bipartite entanglement within the Heisenberg dimers is approximately two-times stronger in the quantum ferrimagnetic phase $|{\rm{II}}\rangle$ than in the quantum antiferromagnetic phase $|{\rm{I}}\rangle$ for this choice of the interaction constants [see Fig. \ref{fig5}(a)]. An increase of temperature causes a gradual smoothing of the concurrence, which is successively suppressed by thermal fluctuations above both quantum ground states $|{\rm{I}}\rangle$ and $|{\rm{II}}\rangle$ and contrarily reinforced above the classical ground state $|{\rm{III}}\rangle$.

Typical temperature dependencies of the concurrence in the same cut of the parameter space $J_1/J=-1.5$ are depicted in Fig. \ref{fig5}(b) for several values of the magnetic  field. It is evident from Fig. \ref{fig5}(b) that the concurrence mostly monotonically decreases with increasing temperature though it may also show a more striking nonmonotonous temperature dependence, specifically slightly below the saturation field. Indeed, the concurrence shows at the saturation field  $h_{c2}/J=2.25$ a gradual thermally-induced decline starting from the zero-temperature asymptotic value $C \approx 0.4$ due to a coexistence of the phases $|{\rm{II}}\rangle$ and  $|{\rm{III}}\rangle$, while it exhibits a vigorous thermally-induced decline (rise) just below (just above) of the saturation field $h/J=2.2$ ($h/J=2.3$) owing to thermal excitations to the classical ferromagnetic (quantum ferrimagnetic) phase $|{\rm{III}}\rangle$ ($|{\rm{II}}\rangle$). 

\begin{figure*}[t]
\begin{center}
\includegraphics[width=0.45\textwidth]{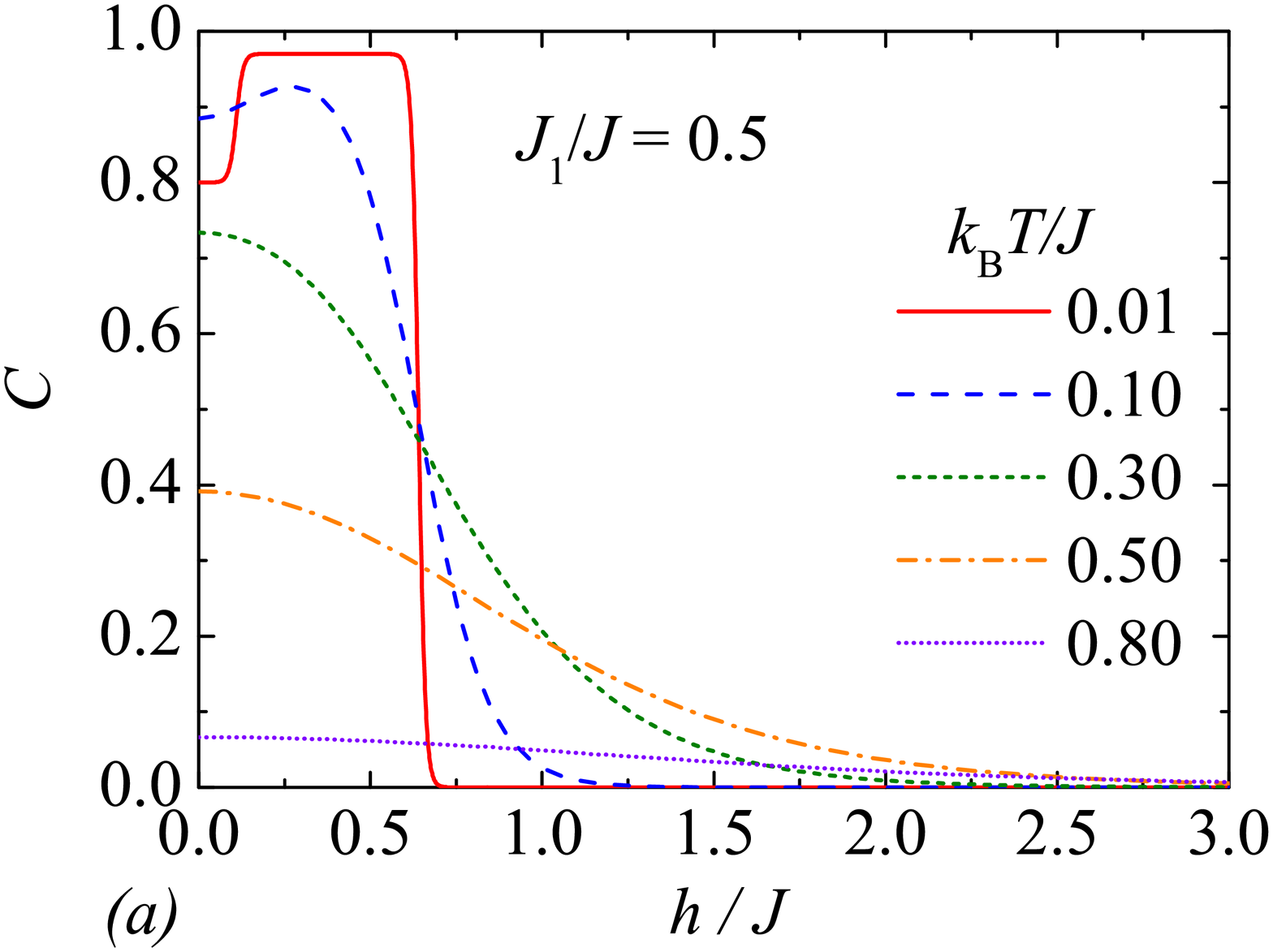}
\hspace{-0.2cm}
\includegraphics[width=0.45\textwidth]{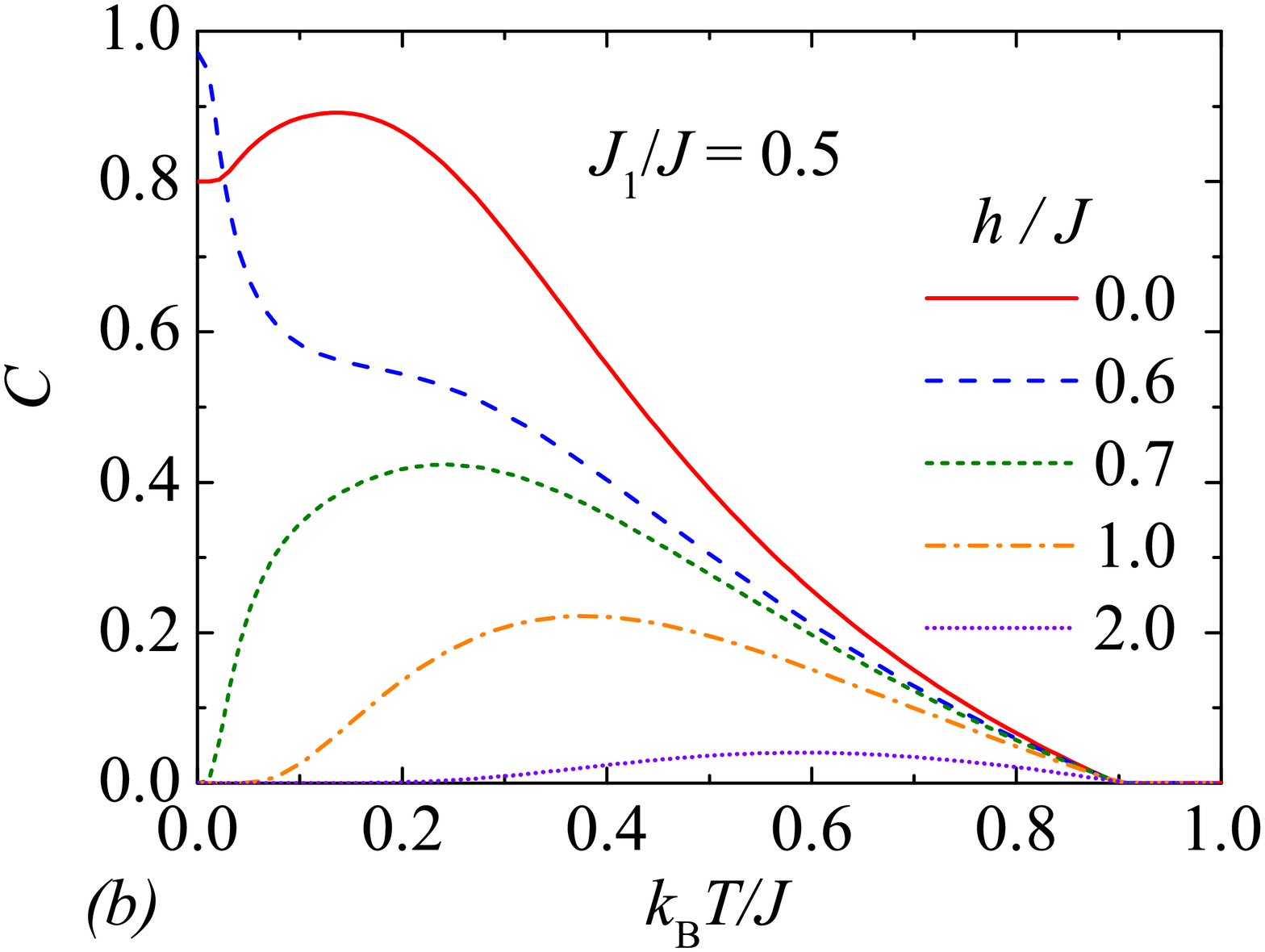}
\hspace{-0.2cm}
\end{center}
\vspace{-0.4cm}
\caption{(a) The concurrence as a function of the magnetic field for the interaction ratio $J_1/J=0.5$ and several values of temperature; (b) Temperature variations of the concurrence for the interaction ratio $J_1/J=0.5$ and several magnetic fields.}
\label{fig6}
\end{figure*}

The concurrence is plotted in Fig. \ref{fig6}(a) against the magnetic field at fixed value of the interaction ratio $J_1/J=0.5$  and a few different temperatures. Under this condition, the concurrence exhibits stepwise changes close to the  critical fields $h_{c1}/J\approx 0.11$ and $h_{c2}/J\approx 0.64$. Apparently, a small rise of temperature can invoke  increase of the concurrence, specifically, the concurrence at zero magnetic field is higher at moderate temperature $k_{\rm{B}}T/J=0.1$ than at very low temperature $k_{\rm{B}}T/J=0.01$ on account of thermal excitations from the less entangled phase $|{\rm{I'}}\rangle$ towards the more entangled phase $|{\rm{II}}\rangle$. To support this statement,  temperature dependence of the concurrence is depicted in Fig. \ref{fig6}(b) for the interaction ratio $J_1/J=0.5$ and several values of the magnetic field. The displayed temperature dependence of the concurrence at zero magnetic field indeed corroborates a transient strengthening of the bipartite entanglement within the range of moderate temperatures ($k_{\rm{B}}T/J \leq 0.2$), which is successively followed by a relatively steep decrease at higher temperatures. 
The concurrence thus starts from its highest possible  value for $J_1/J=0.5$  in a range of moderate  magnetic fields $h/J \in (0.11;0.64)$, which stabilize the phase $|{\rm{II}}\rangle$ in  concordance with the ground-state phase diagram shown in Fig. \ref{fig3}(a).  Unlike this, the concurrence starts from zero above the saturation field in agreement with the classical character of the phase $|{\rm{III}}\rangle$ [c.f. Fig. \ref{fig6}(a)], but afterwards it shows a marked temperature-induced rise until  a  round maximum is reached that is successively followed by a steep  decrease upon increasing of temperature. It is worthwhile to remark that the concurrence approaches zero close to a  threshold temperature  $k_{\rm{B}}T/J\cong 0.9$, above which it equals zero independently of the magnetic field.

\begin{figure*}[t]
\begin{center}
\includegraphics[width=0.45\textwidth]{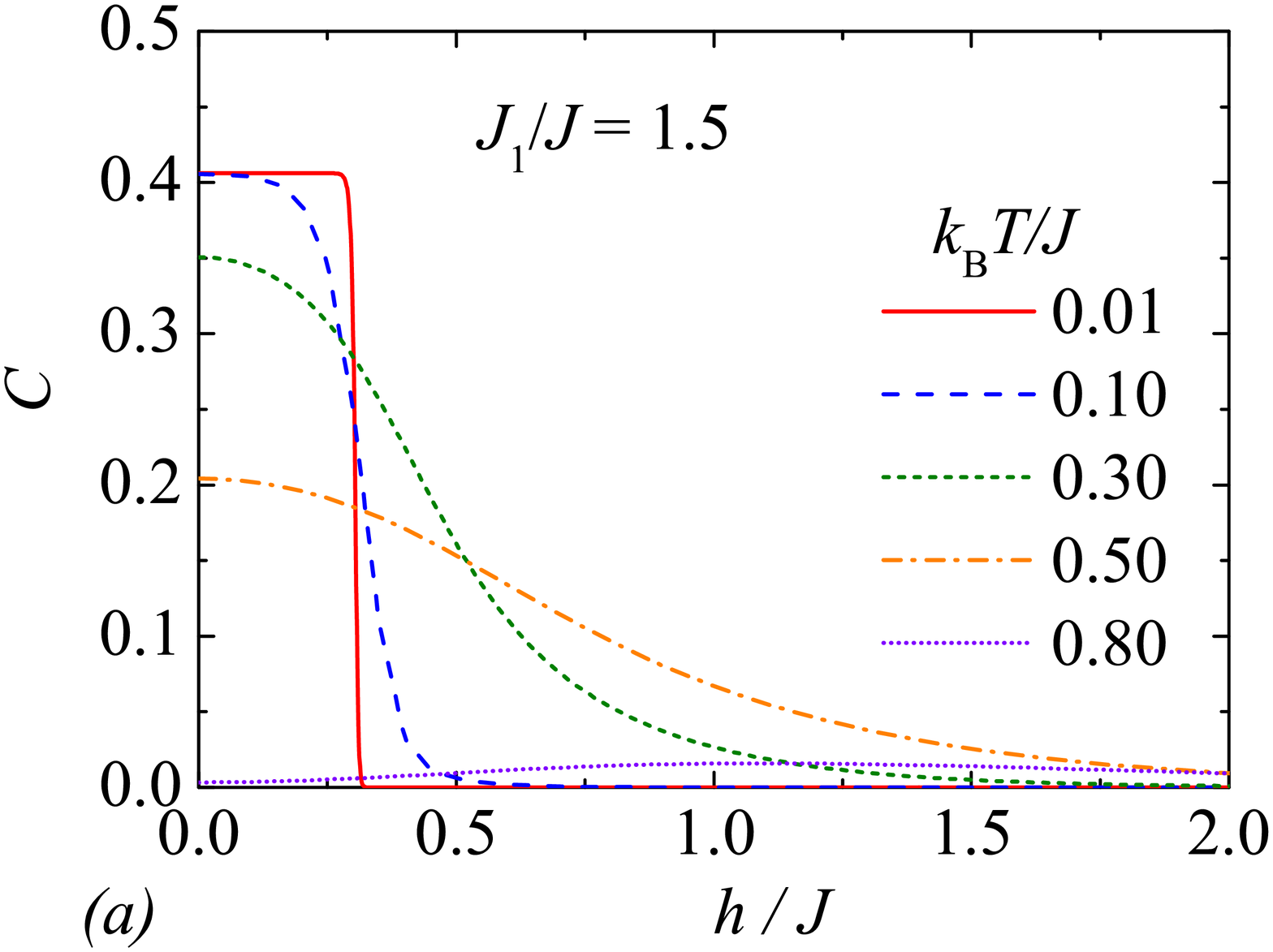}
\hspace{-0.2cm}
\includegraphics[width=0.45\textwidth]{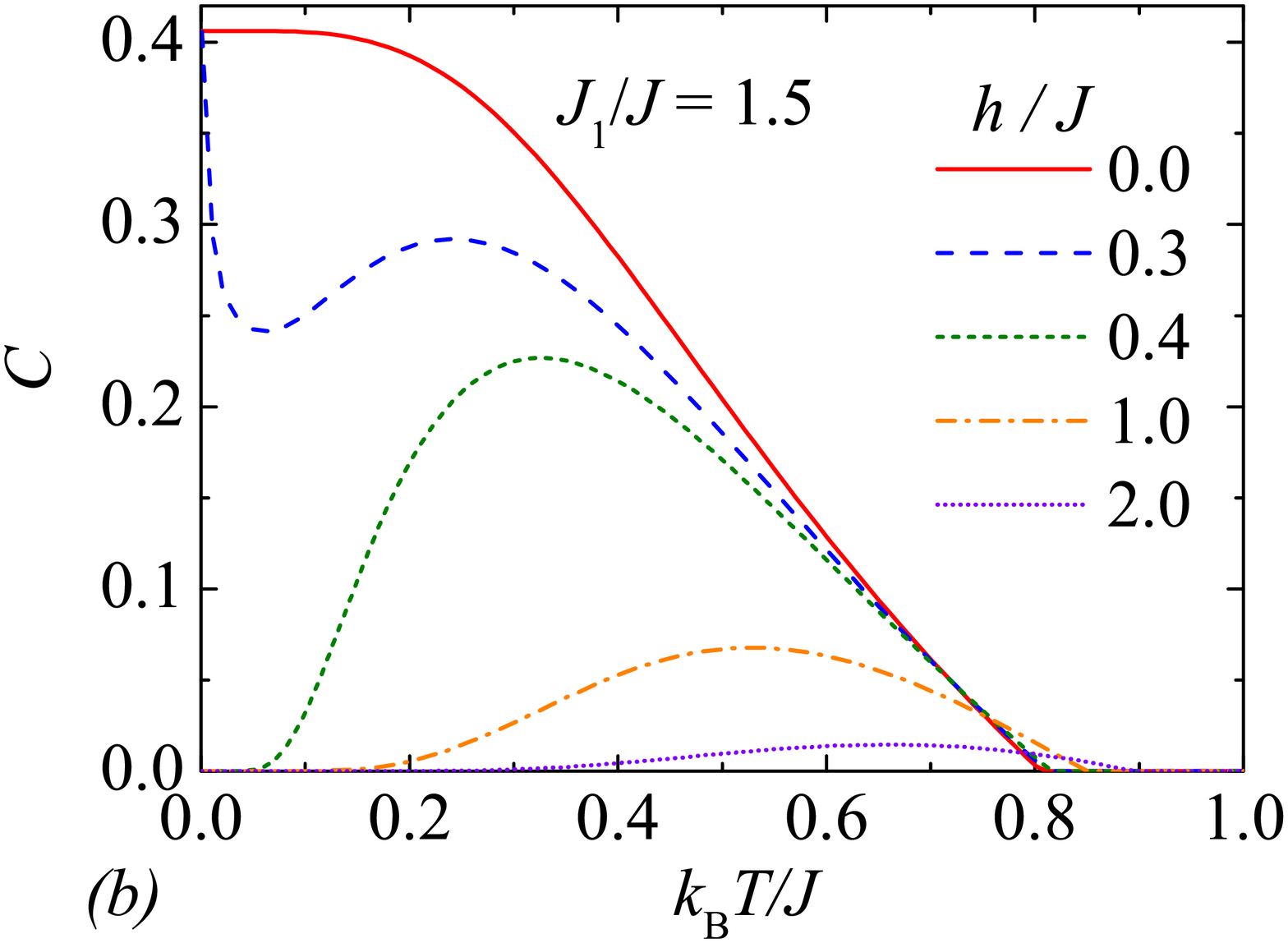}
\hspace{-0.2cm}
\end{center}
\vspace{-0.4cm}
\caption{(a) The concurrence as a function of the magnetic field for the interaction ratio $J_1/J=1.5$ and several values of temperature; (b) Temperature variations of the concurrence for the interaction ratio $J_1/J=1.5$ and several  magnetic fields.}
\label{fig7}
\end{figure*}

Last but not least, we have examined field and temperature dependencies of the concurrence for the interaction ratio $J_1/J=1.5$, which is consistent with only two ground states $|{\rm{I'}}\rangle$ and $|{\rm{III}}\rangle$ in accordance with the established  ground-state phase diagram [see Fig. \ref{fig3}(a)]. In this particular case, one observes an abrupt fall of the concurrence at the saturation field $h_c/J \approx 0.3$, which determines a field-driven phase transition from the phase $|{\rm{I'}}\rangle$ towards the phase $|{\rm{III}}\rangle$. It is apparent from Fig. \ref{fig7}(a) that the concurrence is gradually suppressed upon strengthening of the magnetic field. The most peculiar temperature dependencies of the concurrence can be detected when the magnetic field is selected sufficientlty close but slightly below the saturation field (e.g. $h_c/J = 0.3$ for $J_1/J=1.5$ in Fig. \ref{fig7}(b)). Under this condition, the concurrence shows at very low temperature a steep decline until it reaches a local minimum, then it passes through a round local maximum emergent at moderate temperatures until it finally completely vanishes at the threshold temperature $k_{\rm{B}}T_t/|J_1|=\frac{1}{\ln 3}\approx 0.9$.

\section{Heisenberg branched chain}
\label{H}
Next, let us consider the analogous but purely quantum spin-1/2 Heisenberg branched chain (see Fig. \ref{fig2}), which can be defined through the following Hamiltonian 
\begin{eqnarray} 
\label{hami2}
\hat{\cal{H}}\!\!\!&=&\!\!\! \sum_{i=1}^N \left[J\hat{\bf{S}}_{1,i}\cdot \hat{\bf{S}}_{2,i} -J_1\left(\hat{\bf{S}}_{1,i}\cdot\hat{\bf{S}}_{3,i} + \hat{\bf{S}}_{2,i}\cdot\hat{\bf{S}}_{3,i+1}\right.\right. \nonumber \\
\!\!\!&+&\!\!\!\left.\left.  \hat{\bf{S}}_{1,i}\cdot\hat{\bf{S}}_{4,i}\right) -h\sum_{j=1}^4\hat{S}^z_{j,i}\right].
\end{eqnarray}
Here, $\hat{\bf{S}}_{j,i}$ ($j=1,2,3,4$) are standard spin-1/2 operators assigned to four magnetic ions from the $i$th unit cell, the coupling constant $J>0$ stands for the antiferromagnetic interaction within the dimeric Cu$^{2+}$-Cu$^{2+}$ units of the main chain, the coupling constant $J_1>0$ ($J_1<0$) stands for ferromagnetic (antiferromagnetic) interaction between Cu$^{2+}$ and Fe$^{3+}$ ions (see Fig. \ref{fig1} and \ref{fig2}). The Zeeman's term $h$ accounts for the external magnetic field, $N$ denotes the total number of unit cells and $N_t = 4N$ is the total number of spins. For simplicity, periodic boundary conditions are also assumed.

To obtain the ground-state phase diagram and magnetization process  of the spin-1/2 Heisenberg branched chain (\ref{hami2}) we have performed density-matrix renormalization group (DMRG) simulations  by adapting the subroutine from the Algorithms and Libraries for Physics Simulations (ALPS) project.\cite{bauer} It should be mentioned that the  DMRG data were obtained for the spin-1/2 Heisenberg branched chain with $N=24, 36$ and $48$ unit cells (i.e. with the total number of spins $N_T = 96, 144$ and 192), whereas adequate numerical accuracy was achieved through 12 sweeps at targeted system size when increasing the number of kept states up to 1500 during the final sweeps.

\begin{figure*}[t]
\begin{center}
\includegraphics[width=0.45\textwidth]{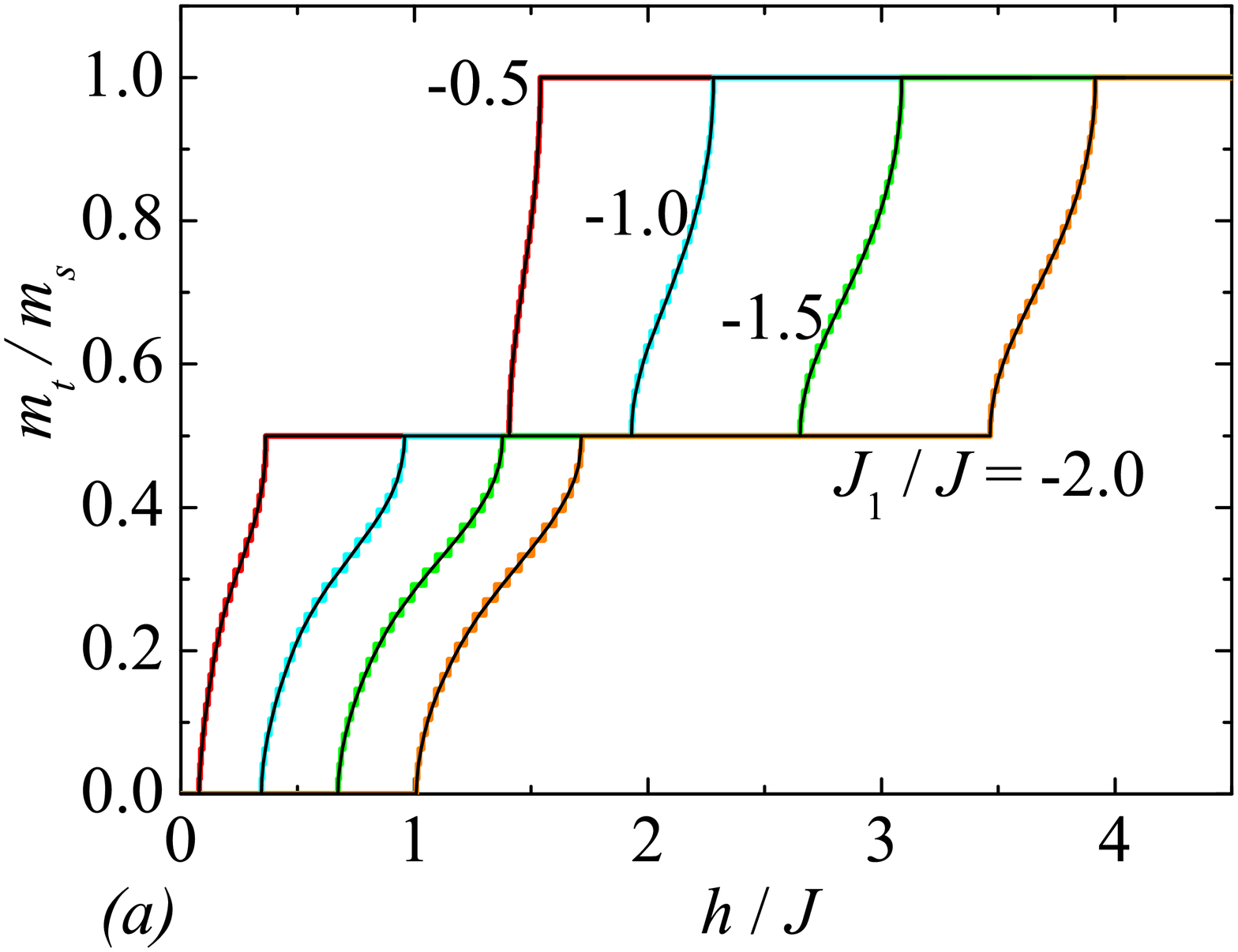}
\hspace{-0.2cm}
\includegraphics[width=0.45\textwidth]{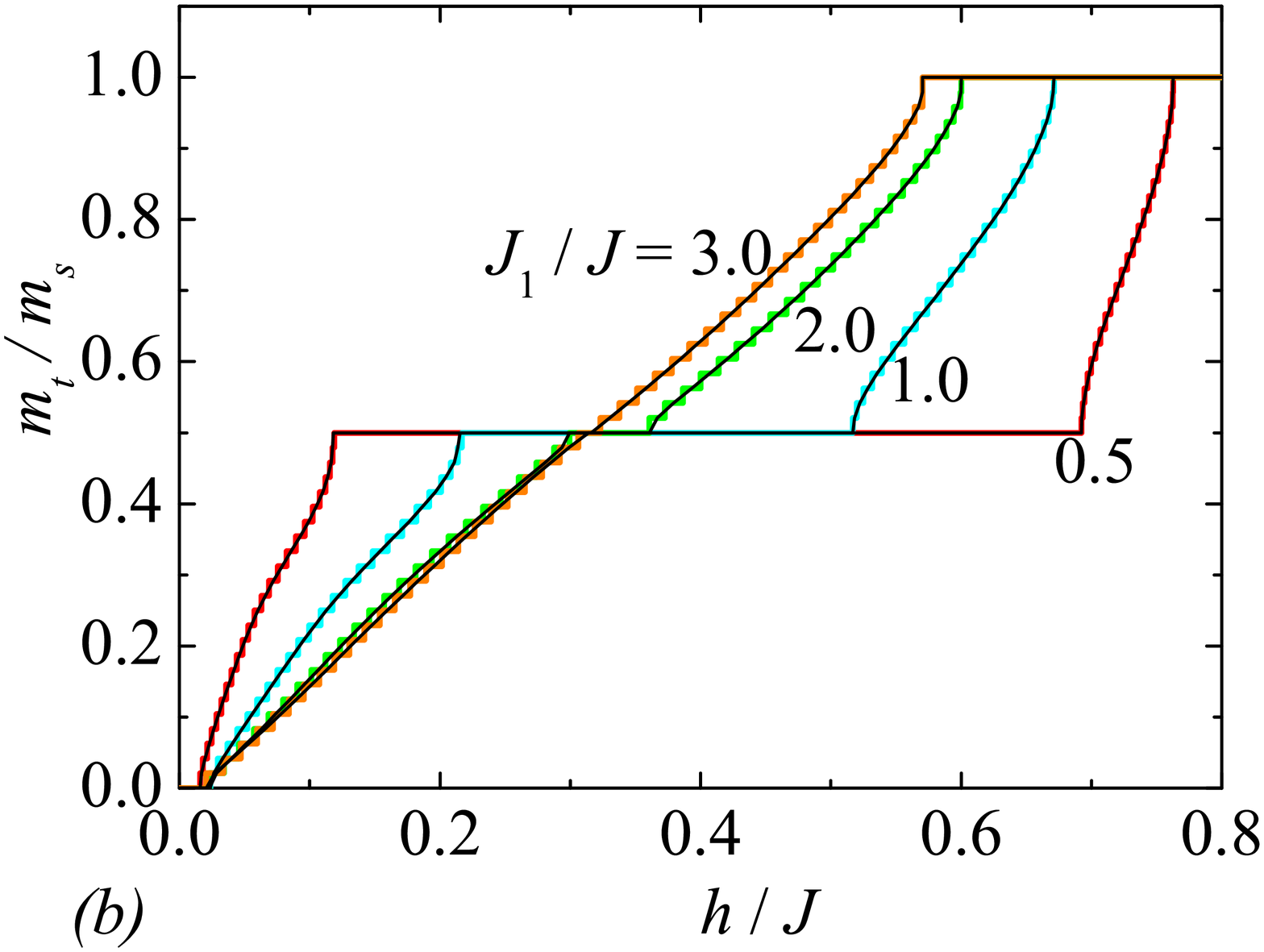}
\hspace{-0.2cm}
\end{center}
\vspace{-0.4cm}
\caption{The magnetic-field dependence of the total magnetization of the spin-1/2 Heisenberg branched chain for several values of the interaction ratio: (a) $J_1/J = -0.5, -1.0, -1.5, -2.0$; (b) $J_1/J=3.0, 2.0, 1.0, 0.5$. Stepwise curves display DMRG data for a finite-size chain with $N=24$ unit cells, while smooth curves are an extrapolation to thermodynamic limit $N \to \infty$.}
\label{fig10}
\end{figure*}

The magnetic-field dependence of the total magnetization is displayed in Fig. \ref{fig10} for several values of interaction ratio $J_1/J$. If the coupling constant $J_1<0$ is antiferromagnetic, the total magnetization of the spin-1/2 Heisenberg branched chain first displays a zero magnetization plateau, which ends up at field-driven quantum phase transition towards a quantum spin liquid terminating at second field-driven quantum phase transition (QPT) towards the one-half magnetization plateau. The intermediate one-half plateau breaks down at third field-driven QPT when the investigated spin chain reenters quantum spin-liquid regime, which terminates at fourth field-driven QPT emergent at a saturation field. For the ferromagnetic coupling constant $J_1>0$, the total magnetization successively exhibits a tiny zero magnetization plateau, quantum spin liquid, one-half magnetization plateau and quantum spin liquid up to a relatively strong ferromagnetic interaction $J_1/J\approx 4$, at which the intermediate one-half plateau vanishes from the magnetization curve. It can be seen from Fig. \ref{fig10} that a width of one-half plateau becomes narrower with increasing of the interaction ratio $J_1/J$  regardless of a sign of the coupling constant $J_1$. Contrary to this, zero magnetization plateau extends over a wider range of the magnetic fields  upon strengthening of the antiferromagnetic coupling constant $J_1<0$, while it  almost remains unchanged with respective of a relative size of  the  ferromagnetic coupling constant $J_1>0$.

\begin{figure*}[t]
\begin{center}
\includegraphics[width=0.45\textwidth]{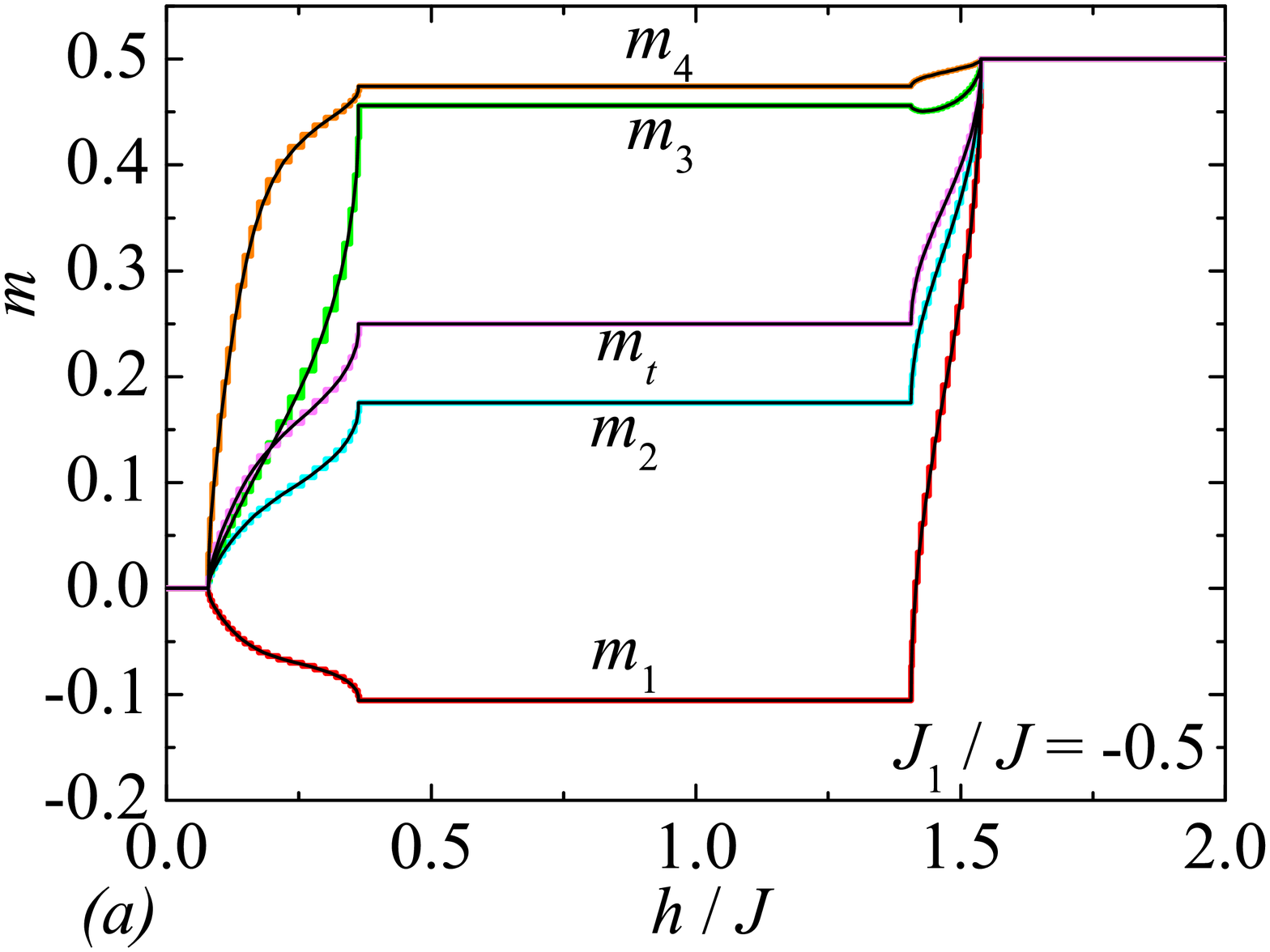}
\hspace{-0.2cm}
\includegraphics[width=0.45\textwidth]{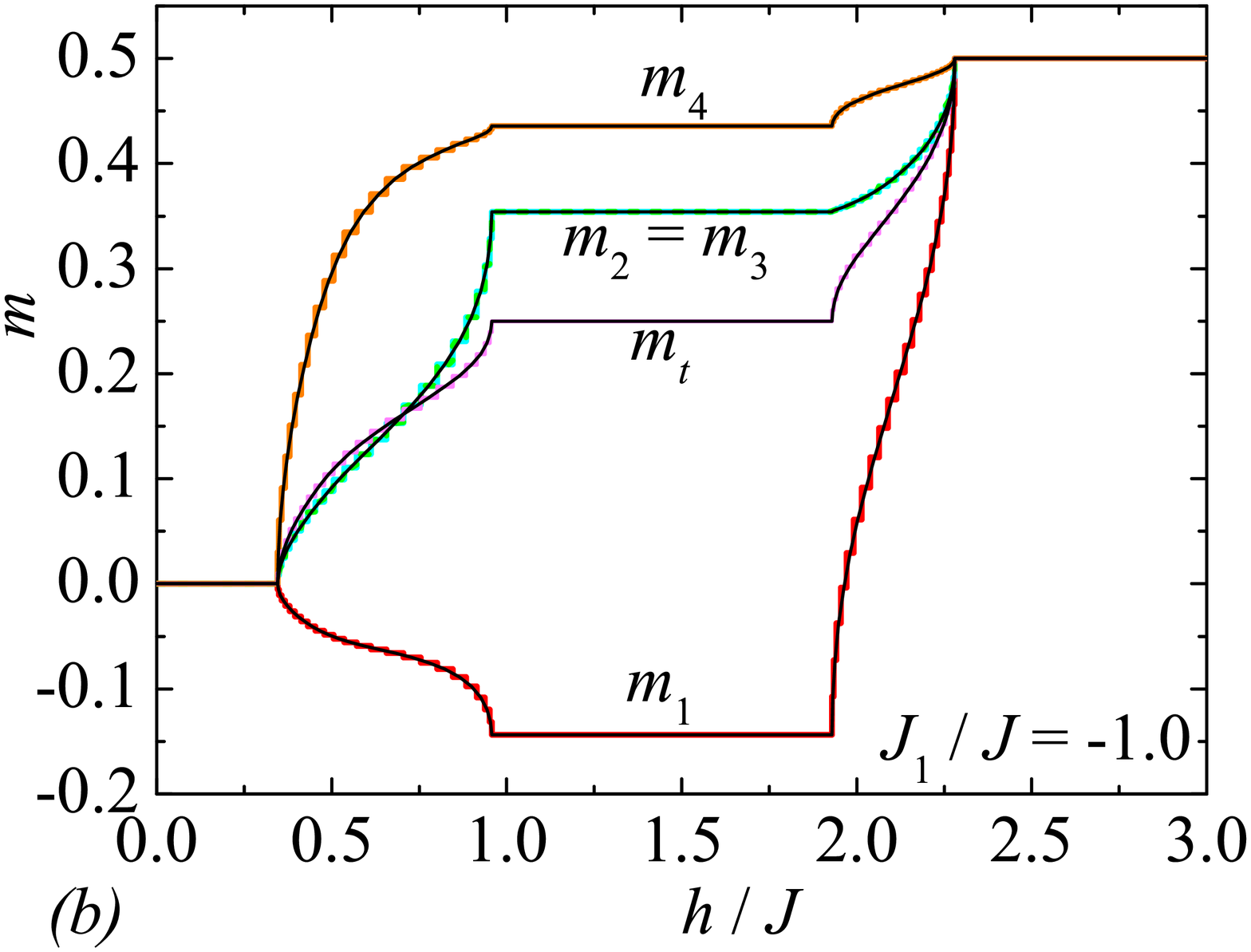}
\hspace{-0.2cm}
\includegraphics[width=0.45\textwidth]{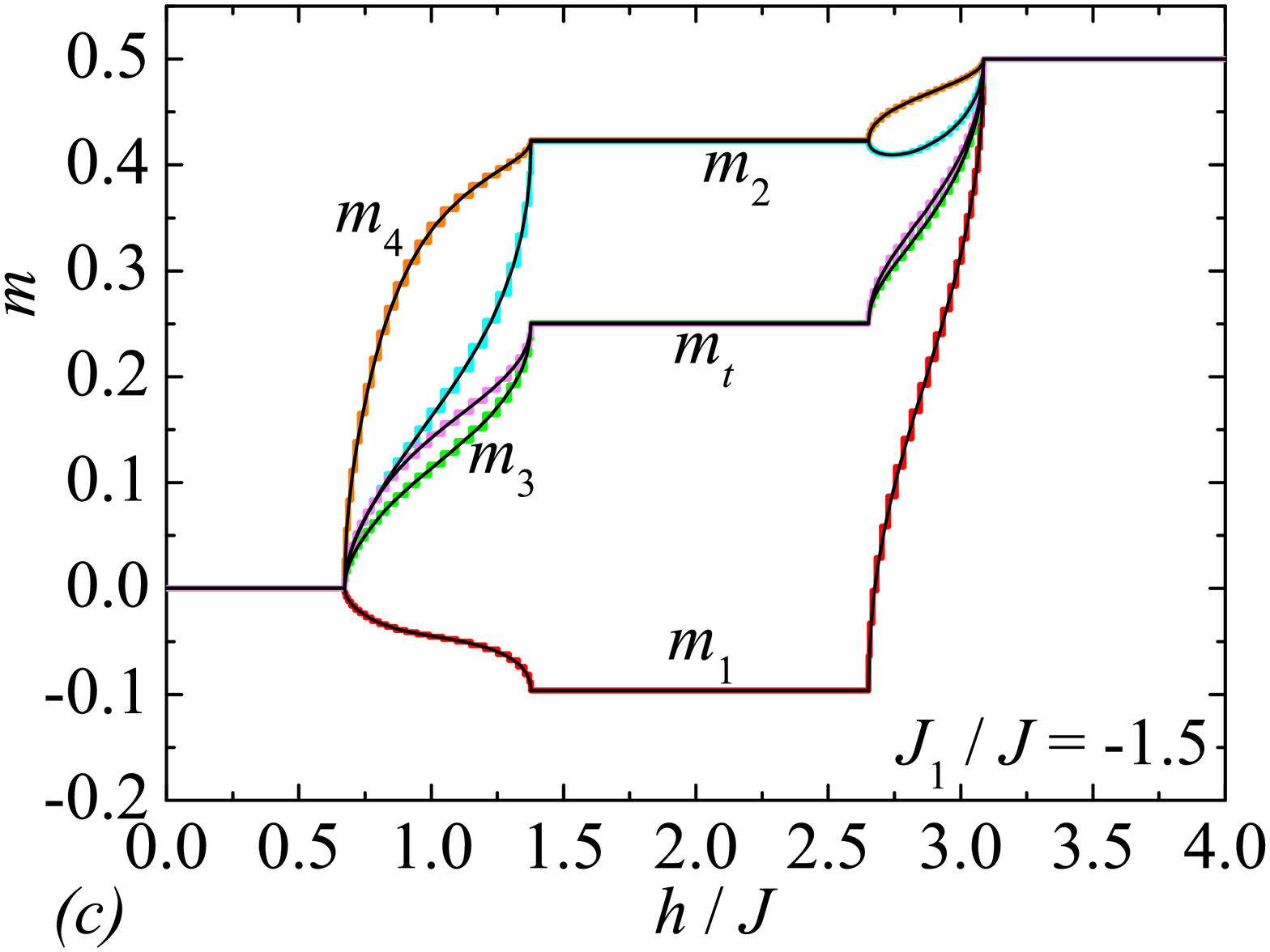}
\hspace{-0.2cm}
\includegraphics[width=0.45\textwidth]{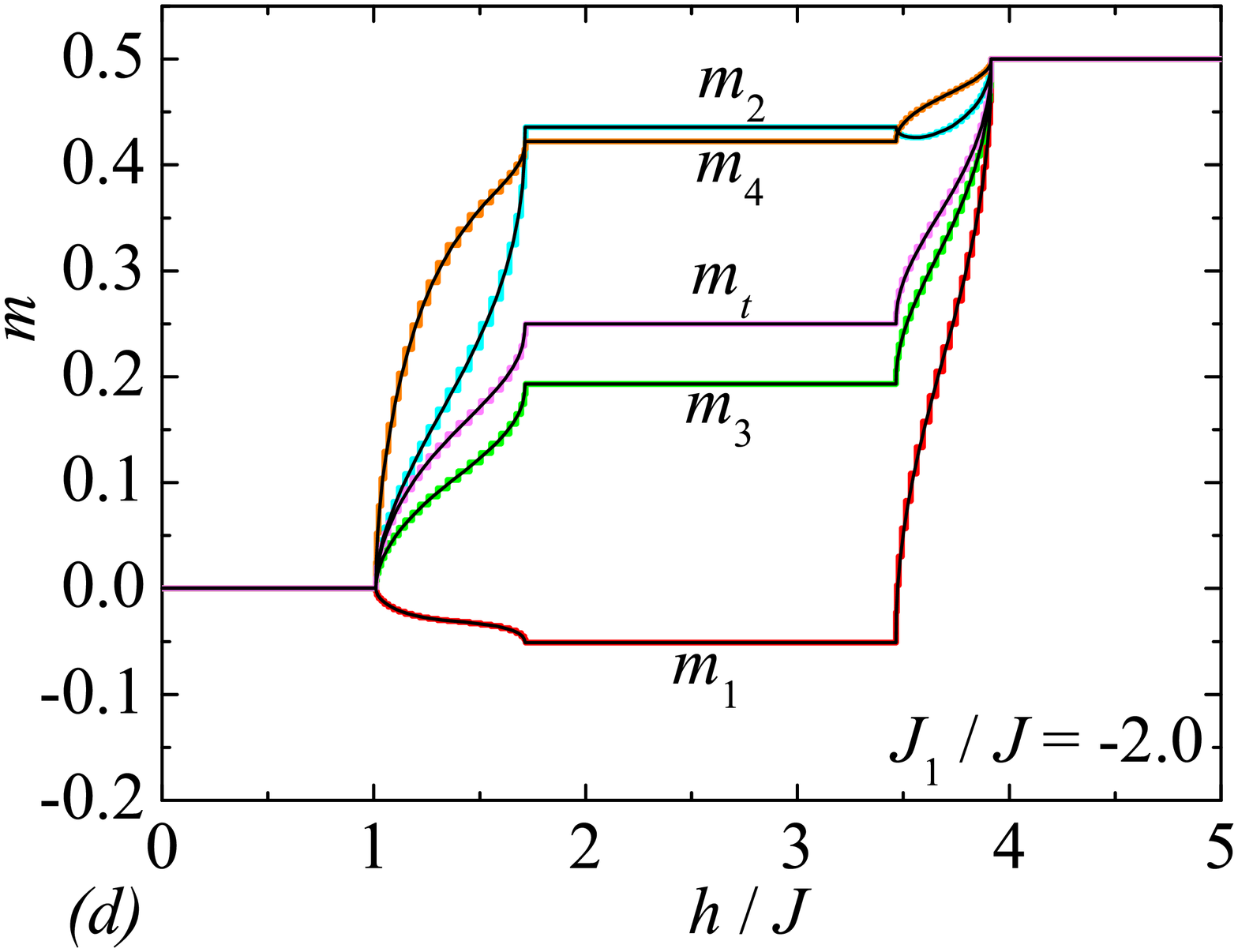}
\hspace{-0.2cm}
\end{center}
\vspace{-0.4cm}
\caption{The magnetic-field dependence of the local and total magnetization of the spin-1/2 Heisenberg branched chain for several values of the interaction ratio:  (a) $J_1/J=-0.5$; (b) $J_1/J=-1.0$; (c) $J_1/J=-1.5$; (d) $J_1/J=-2.0$. Stepwise curves display DMRG data for a finite-size chain with $N=24$ unit cells, while smooth curves were obtained from an extrapolation to thermodynamic limit $N \to \infty$.}
\label{fig8}
\end{figure*}

The magnetic-field dependence of the local and total magnetizations of the spin-1/2 Heisenberg branched chain   are plotted in Fig. \ref{fig8} at zero temperature for several values of the relative size of the antiferromagnetic coupling constant
 $J_1/J$.  It is apparent from Fig. \ref{fig8} that the local magnetizations $m_2, m_3$ and $m_4$ ($\langle S_{2,i}\rangle, \langle S_{3,i}\rangle$ and $\langle S_{4,i}\rangle$) are positive albeit not yet fully saturated in a full range of the magnetic fields, while the local magnetization $m_1$ ($\langle S_{1,i}\rangle$) changes its sign. It should be stressed that the local magnetization $m_4$ achieves within the one-half plateau almost saturated value, while there is evident a quantum reduction in all other local magnetizations due to a presence of quantum fluctuations. It is noteworthy that all local magnetizations generally differ from one another. However,  the local magnetizations may become identical  as for instance the local magnetizations $m_2$ and $m_3$ for the special value of interaction ratio $J_1/J=-1.0$ in a full range of the magnetic fields [see Fig. \ref{fig8}(b)], or as the local magnetizations $m_2$ and $m_4$ for the coupling ratio $J_1/J=-1.5$ within the one-half plateau [see Fig. \ref{fig8}(c)].

\begin{figure*}[t]
\begin{center}
\includegraphics[width=0.45\textwidth]{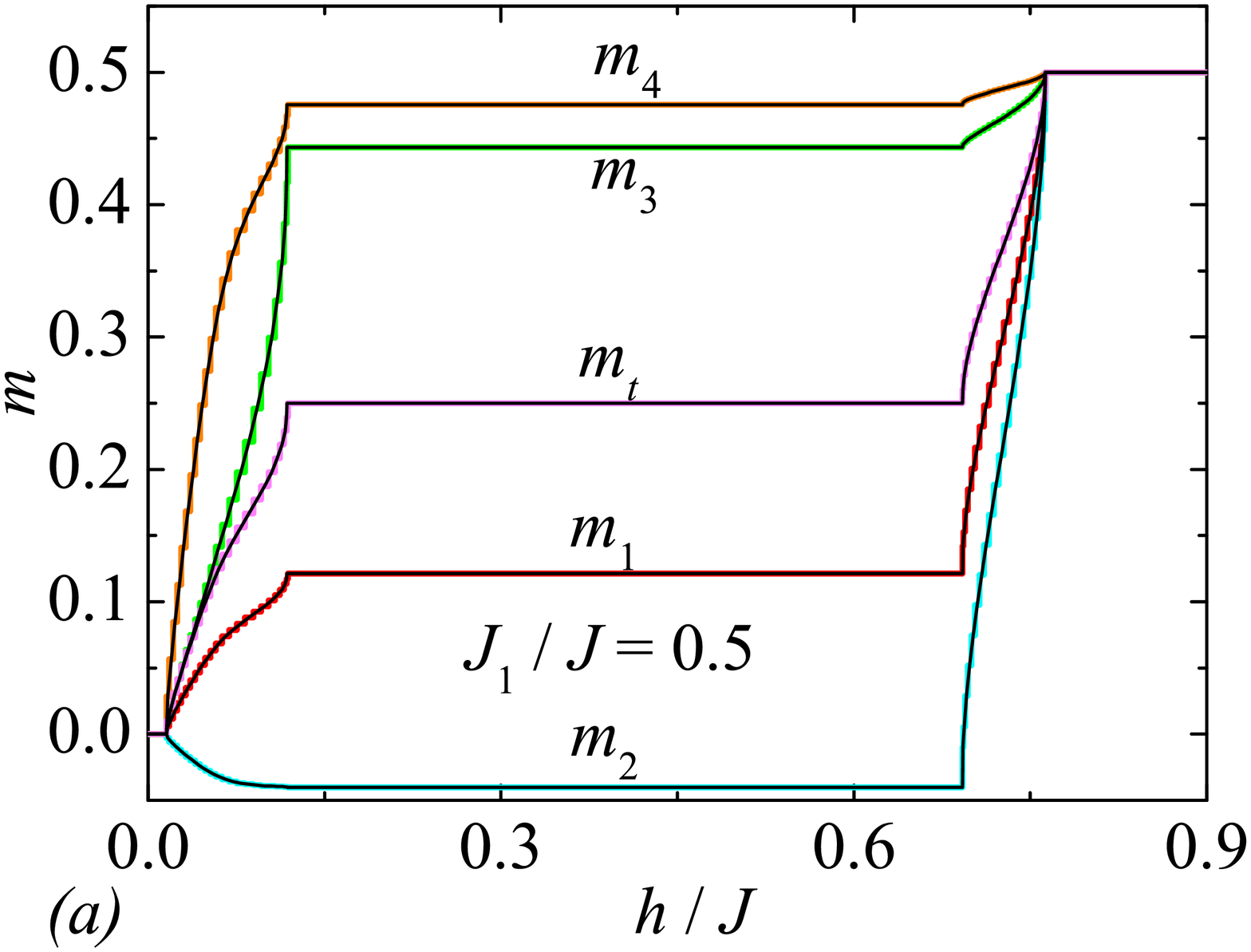}
\hspace{-0.2cm}
\includegraphics[width=0.45\textwidth]{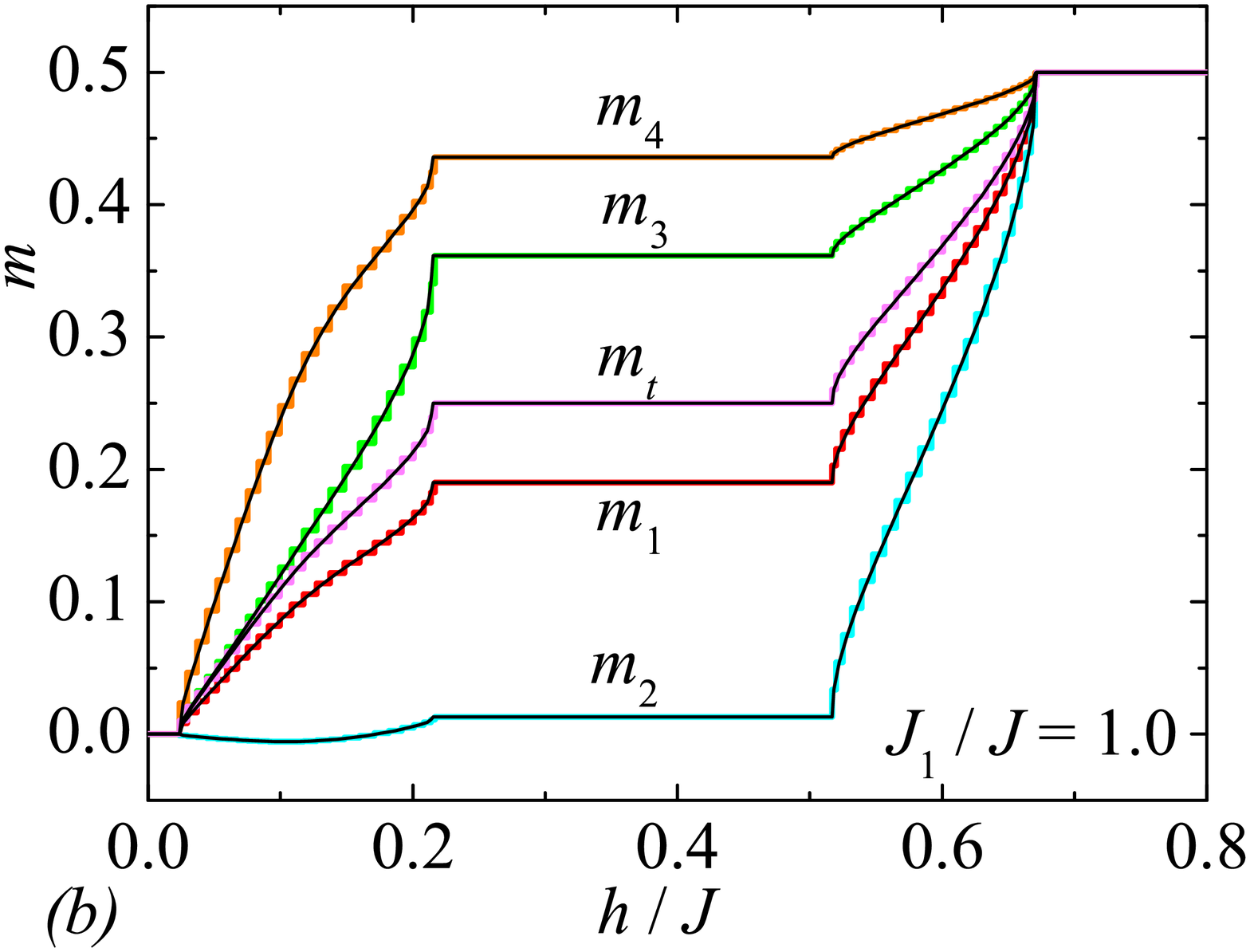}
\hspace{-0.2cm}
\includegraphics[width=0.45\textwidth]{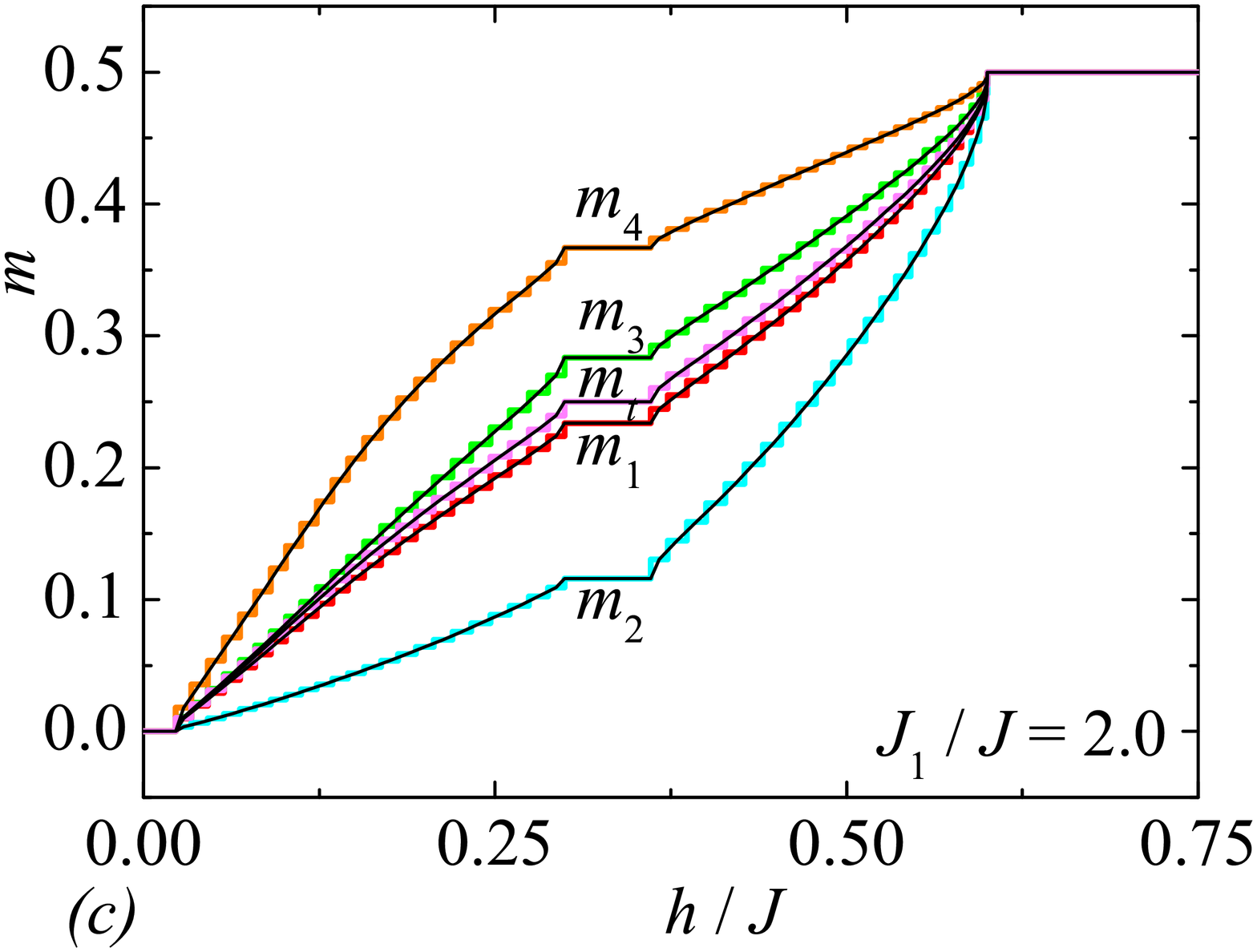}
\hspace{-0.2cm}
\includegraphics[width=0.45\textwidth]{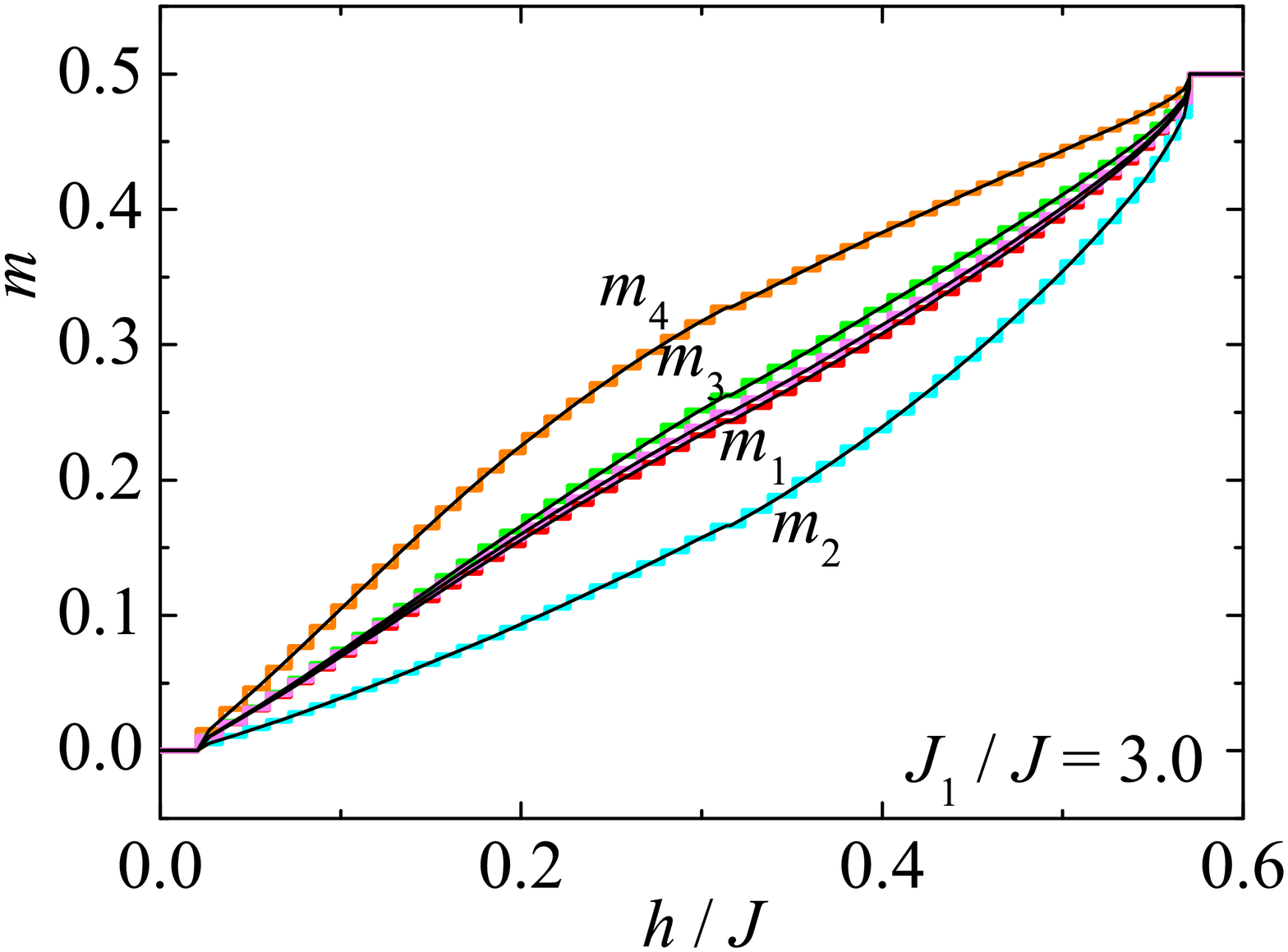}
\hspace{-0.2cm}
\end{center}
\vspace{-0.4cm}
\caption{The magnetic-field dependence of the local and total magnetization of the spin-1/2 Heisenberg branched chain for several values of the interaction ratio:  (a) $J_1/J=-0.5$; (b) $J_1/J=1.0$; (c) $J_1/J=2.0$; (d) $J_1/J=3.0$. Stepwise curves display DMRG data for a finite-size chain with $N=24$ unit cells, while smooth curves are an extrapolation to thermodynamic limit $N \to \infty$.}
\label{fig9}
\end{figure*}

The magnetic-field dependencies of local and total magnetizations for the spin-1/2 Heisenberg branched chain with ferromagnetic coupling constant $J_1>0$ are depicted in Fig. \ref{fig9}. It turns out that all local magnetizations are positive in a full range of the magnetic fields except the local magnetization $m_2$, which may be oriented in opposite to the external magnetic field (negative) [see Fig. \ref{fig9}(a) and (b)]. It is quite curious that the local magnetization $m_4$ is almost saturated at weak ferromagnetic coupling constant $J_1/J \gtrsim 0$ [see Fig. \ref{fig9}(a) and Fig. \ref{fig9}(b)], which means that it does not principally matter whether the spins $S_{4,i}$ are described by the notion of classically Ising or fully quantum Heisenberg spins. Hence, it follows that the spin-1/2 Heisenberg branched chain should resemble in this particular limit a magnetic behavior of the spin-1/2 Ising-Heisenberg branched chain. It should be pointed out, moreover, that the one-half magnetization plateau of the spin-1/2 Heisenberg branched chain generally shrinks upon strengthening of the interaction ratio $J_2/J_1$. Owing to this fact, the one-half plateau emerges just within a very narrow range of the magnetic fields and it becomes almost indiscernible within the used scale as depicted for instance in Fig. \ref{fig9}(d) for the interaction ratio $J_1/J=3$.

\begin{figure*}[t]
\begin{center}
\includegraphics[width=0.45\textwidth]{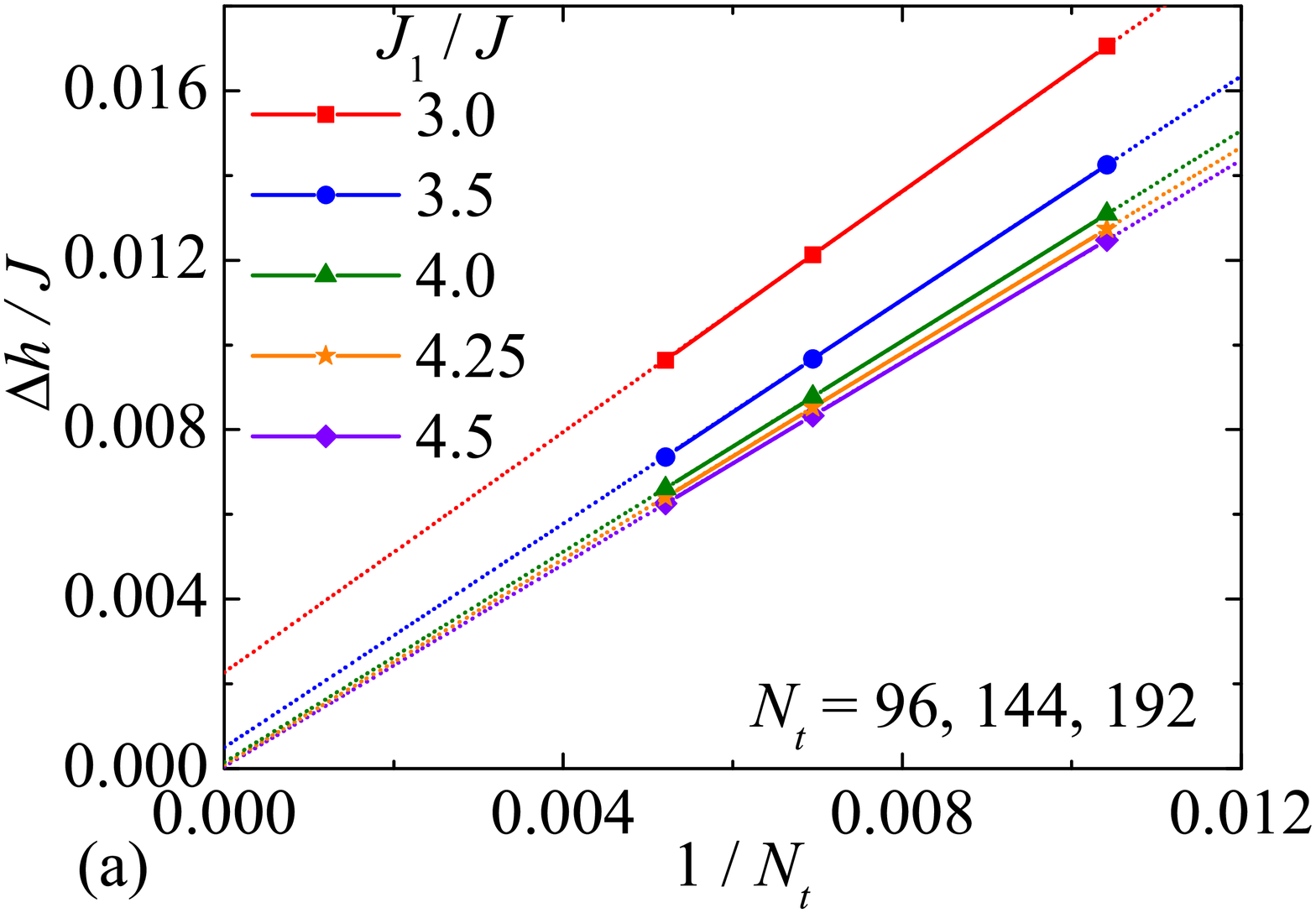}
\hspace{-0.2cm}
\includegraphics[width=0.45\textwidth]{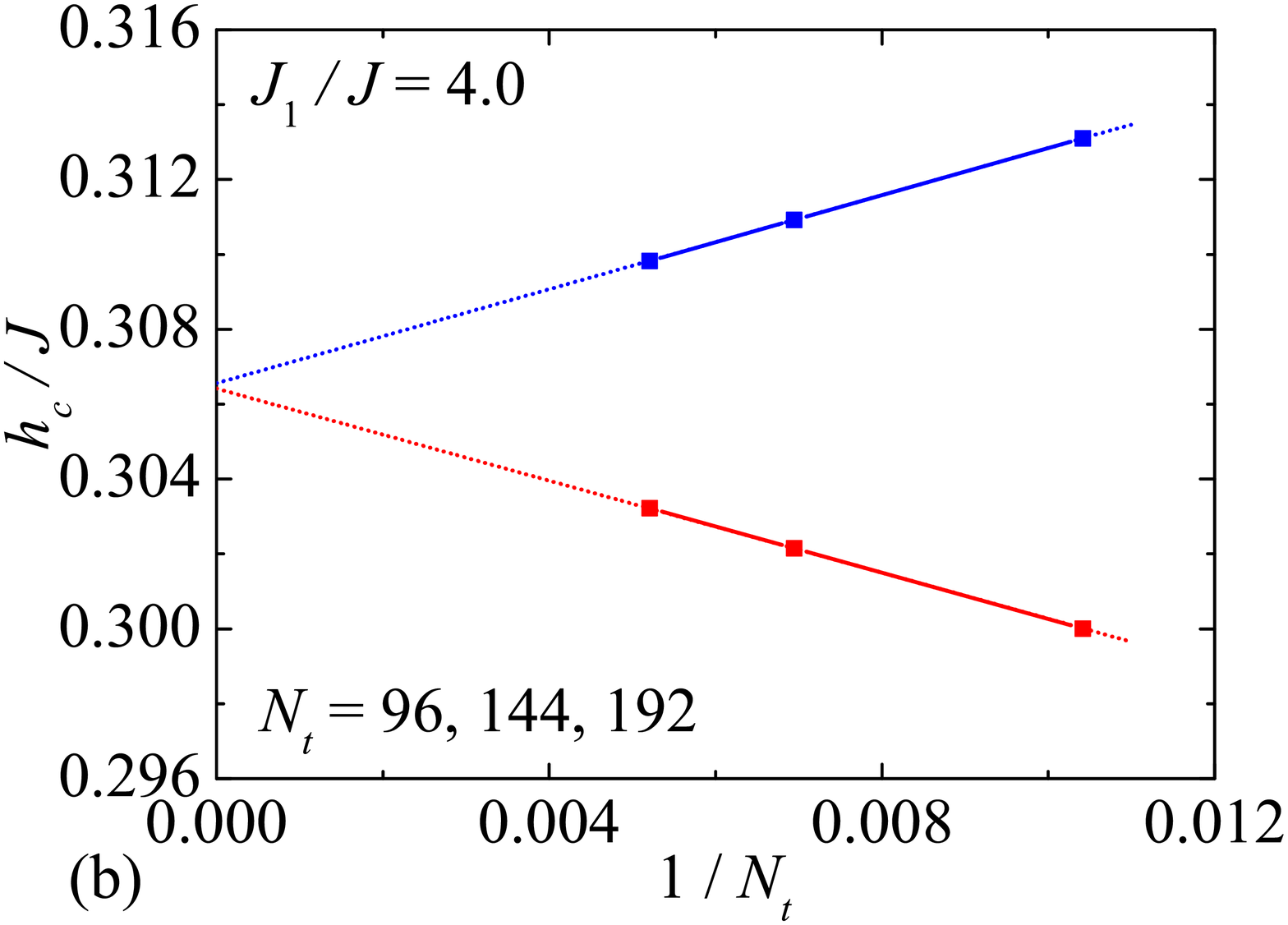}
\vspace{-0.3cm}
\includegraphics[width=0.45\textwidth]{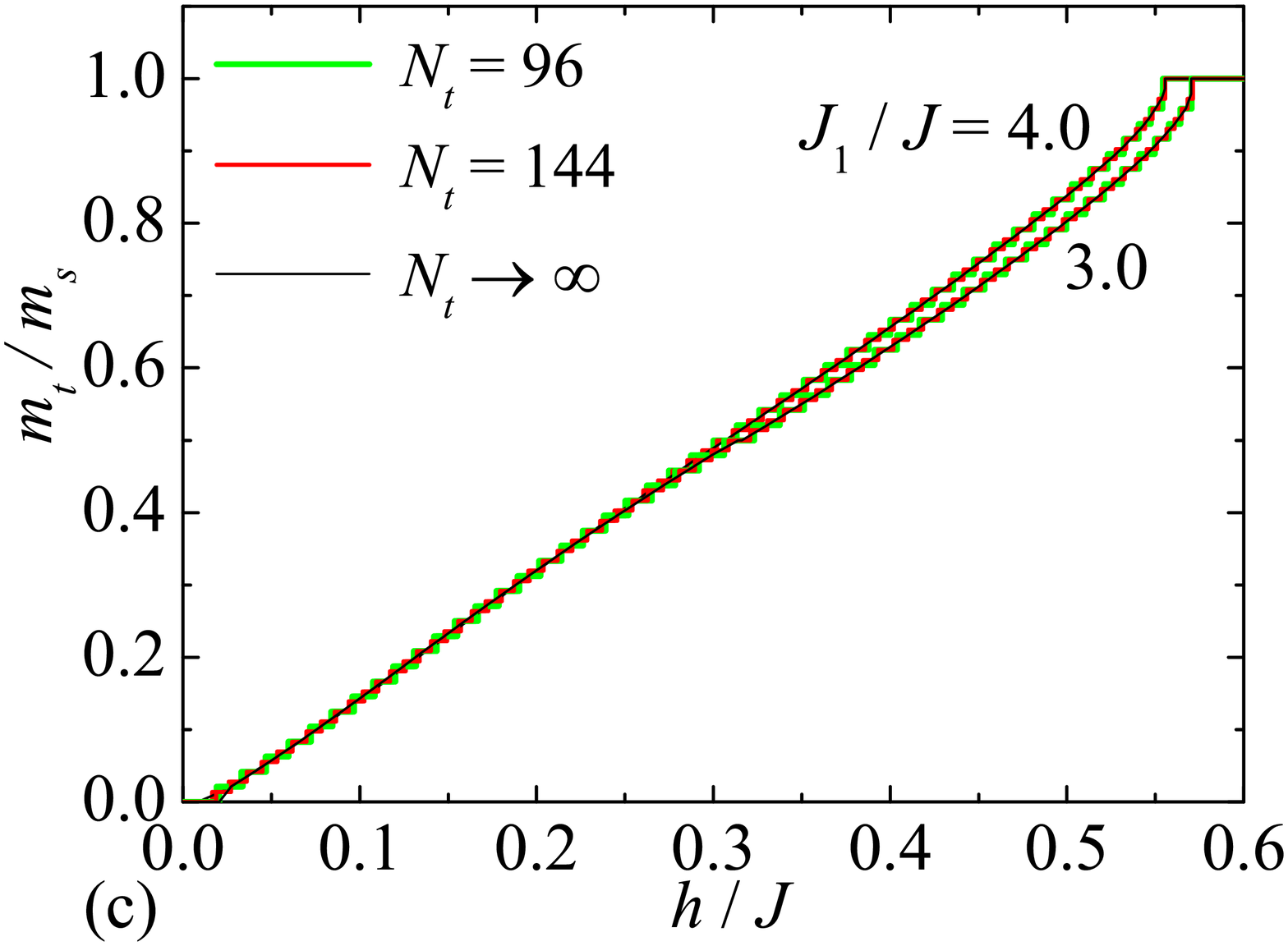}
\hspace{-0.2cm}
\includegraphics[width=0.45\textwidth]{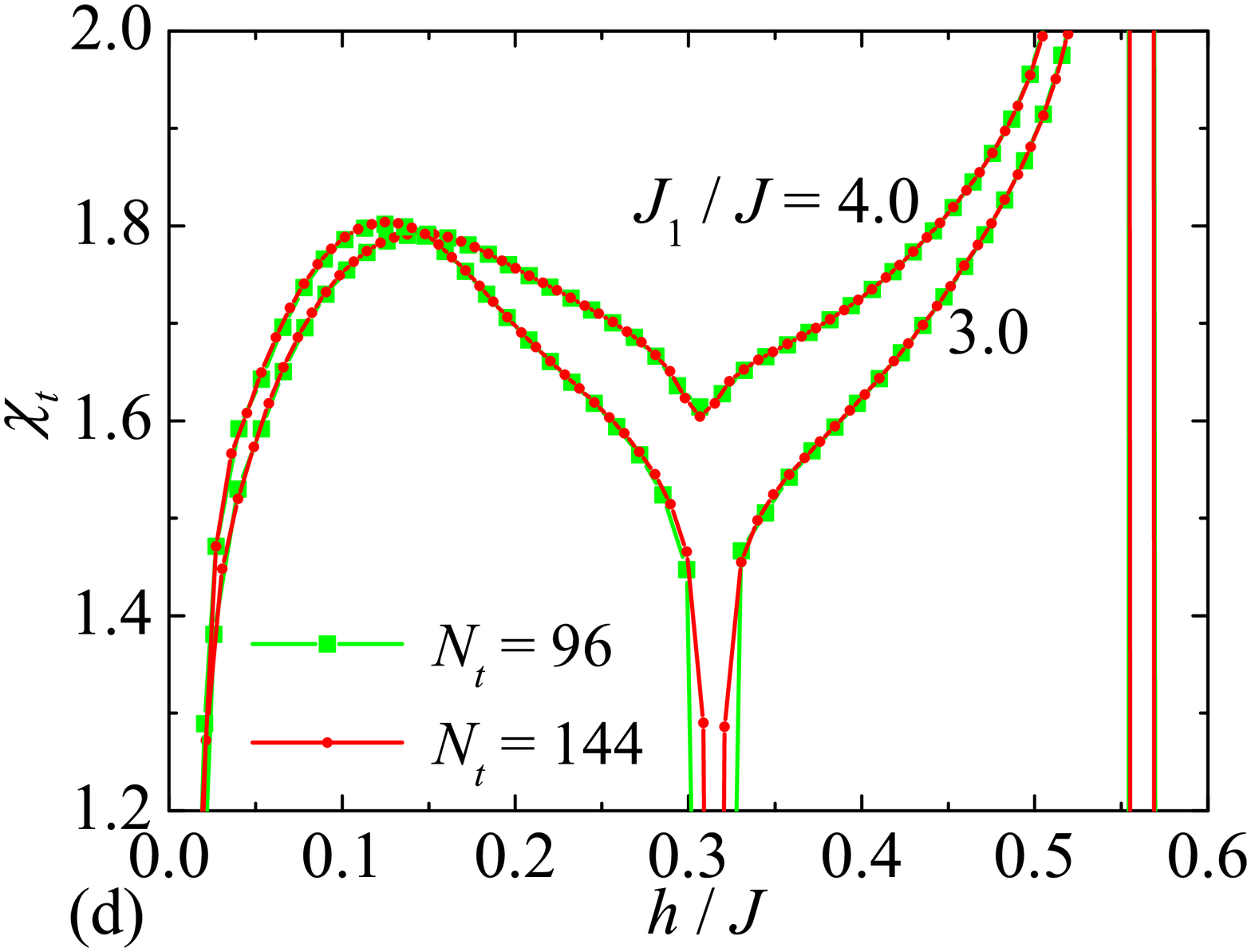}
\hspace{-0.2cm}
\end{center}
\vspace{-0.4cm}
\caption{(a) A width of the intermediate one-half plateau versus a reciprocal value of the total number of spins $N_t$, for a few different values of the interaction ratio $J_1/J$; (b) Upper and lower magnetic fields of the magnetization sector, which might be responsible for intermediate one-half plateau against a reciprocal value of the total number of spins $N_t$ for the interaction ratio $J_1/J = 4.0$; (c) The total magnetization as a function of the magnetic field for the interaction ratios  $J_1/J = 3.0$ and $J_1/J = 4.0$. Stepwise curves are magnetization data obtained from DMRG simulation of  finite-size chains with the total number of spins $N_t = 96$ and 144 (i.e. $N=24$ and $36$ unit cells), while  smooth curves are  an extrapolation to thermodynamic limit $N_t \to \infty$; (d) The susceptibility as a function of the magnetic field for the interaction ratios  $J_1/J = 3.0$ and $J_1/J = 4.0$ as obtained from DMRG simulation of  finite-size chains with the total number of spins $N_t = 96$ and 144 (i.e. $N=24$ and $36$ unit cells).}
\label{fig11}
\end{figure*}

Let us examine a breakdown of the intermediate one-half magnetization plateau of the spin-1/2 Heisenberg branched chain in a somewhat more detail. A width of the magnetization sector, which corresponds to the intermediate one-half plateau, is plotted in Fig. \ref{fig11}(a) against a reciprocal value of the total number of spins $N_t$ for a few different values of the interaction ratio $J_1/J$. It is quite obvious from Fig. \ref{fig11}(a) that a spin gap associated with existence of the intermediate one-half plateau closes only very gradually upon strengthening of the interaction ratio $J_1/J$. A proper finite-size analysis implies that the one-half magnetization plateau (spin gap) still persists in the thermodynamic limit for the interaction ratio $J_1/J = 3.0$ and $3.5$, while it completely vanishes above a quantum critical point emergent close the interaction ratio $J_1/J \lesssim 4.0$ [see Fig. \ref{fig11}(a)]. The data extrapolated for upper and lower critical fields of the intermediate one-half plateau, which are displayed in Fig. \ref{fig11}(b) for the interaction ratio $J_1/J = 4.0$, are in accordance with this statement. Moreover, an exponentially slow suppression of a spin gap suggests that the intermediate one-half magnetization plateau terminates at a quantum critical point of Kosterlitz-Thouless type quite similarly as recently reported for a quantum critical point of the mixed spin-(1/2,5/2,1/2) Heisenberg branched chain.\cite{veri19}
While the magnetization curve does not bear any clear evidence of this type of quantum criticality [see Fig. \ref{fig11}(c)], the susceptibility should display a pronounced dip at the relevant quantum critical point and zero value within the magnetic-field range corresponding to the intermediate one-half plateau [see Fig. \ref{fig11}(d)]. 
 
\begin{figure*}[t]
\begin{center}
\includegraphics[width=0.45\textwidth]{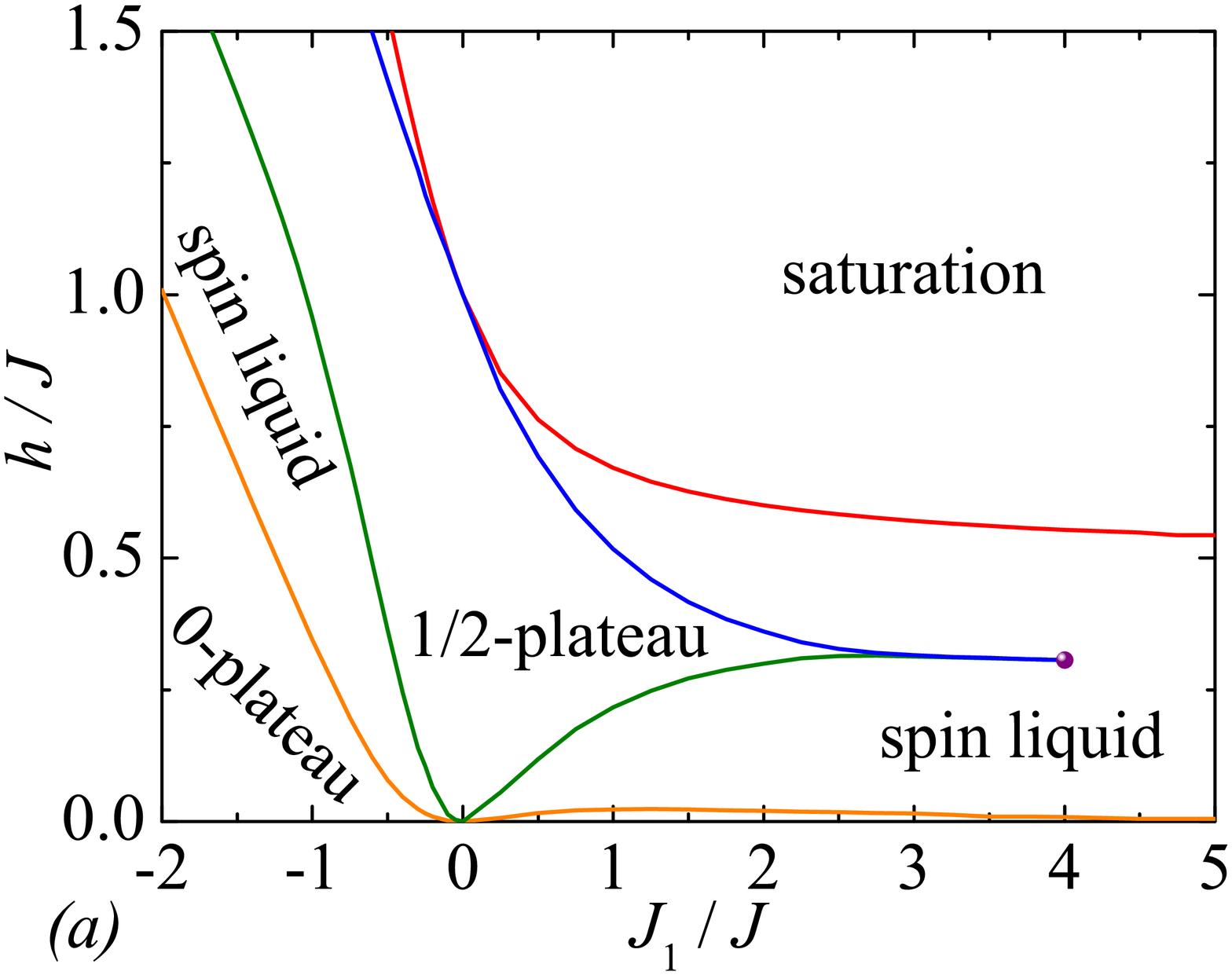}
\hspace{-0.2cm}
\includegraphics[width=0.45\textwidth]{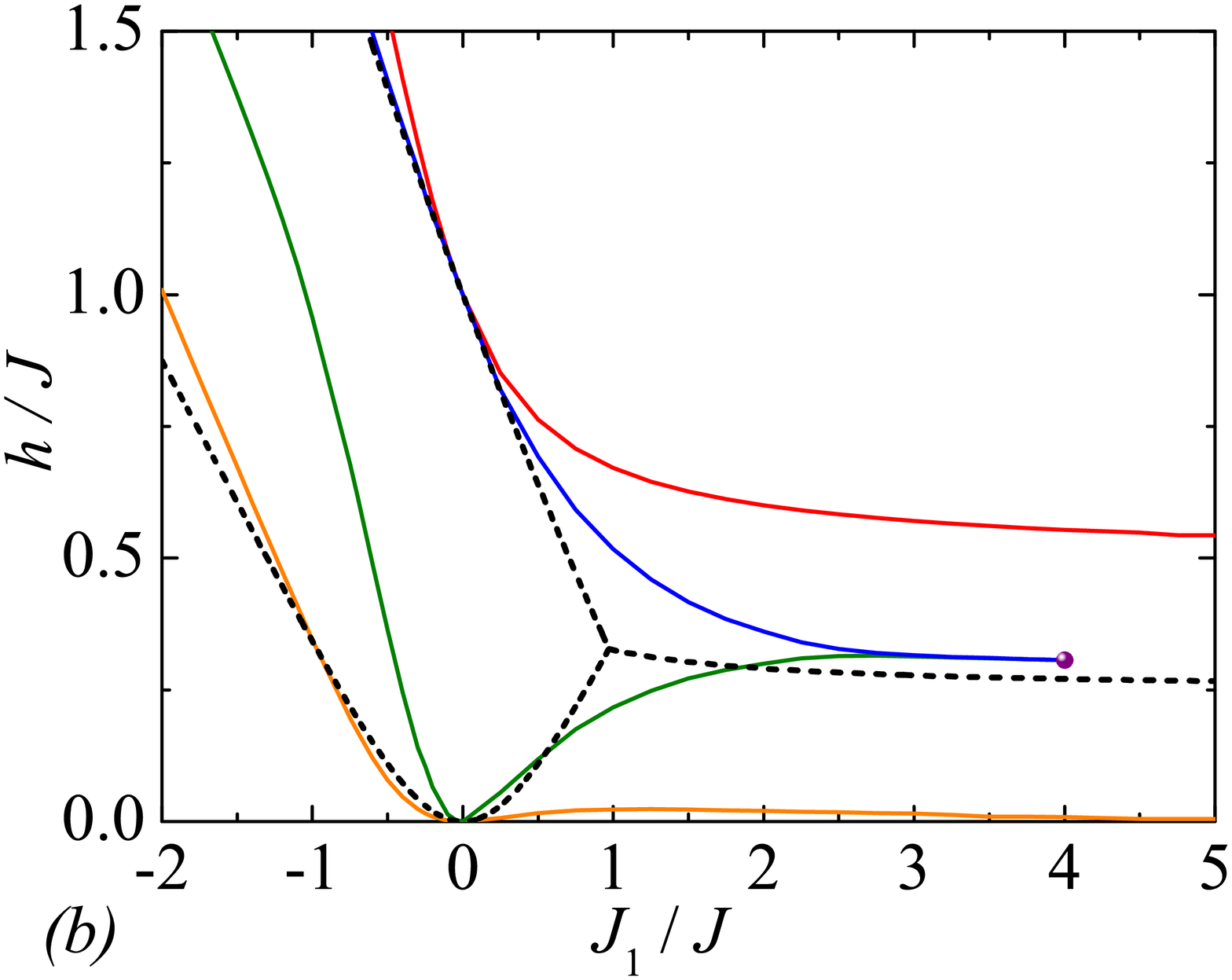}
\hspace{-0.2cm}
\end{center}
\vspace{-0.4cm}
\caption{(a) The ground-state phase diagram of the spin-1/2 Heisenberg branched chain in the $h/J-J_1/J$ plane; (b) A comparison between the ground-state phase diagrams of the spin-1/2 Heisenberg branched chain (solid lines) and the spin-1/2 Ising-Heisenberg branched chain (broken lines).}
\label{fig12}
\end{figure*}

To confirm all aforedescribed results we have plotted in Fig. \ref{fig12}(a) the ground-state phase diagram of the spin-1/2 Heisenberg branched chain in the plane $J_1/J-h/J$. The ground-state phase diagram totally involves zero and one-half plateau, quantum spin liquid and fully polarized ferromagnetic phase. The ground state related to zero plateau becomes narrower upon weakening of the antiferromagnetic coupling constant $J_1<0$, while it holds very low (but nonzero) value for all positive values of the ferromagnetic coupling constant $J_1>0$. Contrary to this, the gapped phase related to the one-half magnetization plateau becomes narrower upon strengthening of the ferromagnetic coupling constant until it completely vanishes above a quantum critical point located around the interaction ratio $J_1/J \lesssim 4.0$, at which gapful one-half plateau phase coexist together with the gapless spin-liquid phase.

Finally, we have plotted in Fig. \ref{fig12}(b) the ground-state phase diagrams of the spin-1/2 Heisenberg branched chain and the spin-1/2 Ising-Heisenberg branched chain in order to compare phase boundaries of both studied systems. It follows from Fig. \ref{fig12}(b) that the intermediate one-half magnetization plateau of the spin-1/2 Ising-Heisenberg branched chain is suppressed by a quantum spin liquid. Moreover, the phase related to zero magnetization plateau of the spin-1/2 Heisenberg branched chain is realized only at very low magnetic fields in comparison with zero magnetization plateau of the spin-1/2 Ising-Heisenberg branched chain. On the other hand, the phase transition between one-half plateau and saturation is achieved almost at the same magnetic field for both models within the interval of interaction ratio $J_1/J<0.25$. The most significant discrepancy is that the intermediate one-half plateau of the spin-1/2 Ising-Heisenberg branched chain ends up at a triple point ($J_2/J \approx 0.97$), while the one-half  plateau of the spin-1/2 Heisenberg branched chain diminishes at the Kosterlitz-Thouless quantum critical point ($J_1/J \lesssim 4.0$), at which phases related to the one-half plateau and the quantum spin liquid coexist together.
\section{Conclusion}
\label{conclusion}
In the present work, we have examined the ground-state phase diagram, magnetization curves and concurrence of the spin-1/2 Ising Heisenberg branched chain by the use of the transfer-matrix method. Besides this, we have provided the numerical simulation in order to obtain the ground-state phase diagram, local and total magnetization of the spin-1/2 Heisenberg branched chain. We have found three different ground states in the spin-1/2 Ising-Heisenberg branched chain depending on a mutual interplay between the magnetic field and two different coupling constants. The modulated quantum antiferromagnetic phase manifests itself in a zero-temperature magnetization process as zero plateau, the quantum ferrimagnetic phase as the intermediate one-half plateau and the classical ferromagnetic phase is trivial fully polarized state. Besides zero and one-half plateau we have discovered a gapless quantum spin-liquid phase in a magnetization process of the spin-1/2 Heisenberg branched chain. The most interesting finding presented in this paper was that while the one-half plateau of the spin-1/2 Ising-Heisenberg branched chain terminates at a triple point, the one-half plateau of the spin-1/2 Heisenberg branched chain ends up at Kosterlitz-Thouless quantum critical point.

\begin{acknowledgments}
This work was financially supported by the grant of The Ministry of Education, Science, Research and Sport of the Slovak Republic under the contract No. VEGA 1/0043/16 and by the grant of the Slovak Research and Development Agency under the contract Nos. APVV-18-0197 and APVV-16-0186. 
\end{acknowledgments}

\end{document}